\documentclass[10pt,journal,final]{IEEEtran}
\usepackage{lineno,hyperref}
\usepackage{balance}
\usepackage{graphicx}
\usepackage{subfigure}
\usepackage{epstopdf}
\usepackage{mathrsfs}
\usepackage{graphicx}
\usepackage{amssymb}
\usepackage{amsmath}
\usepackage{amsfonts}
\usepackage{amsthm}
\usepackage{color}
\usepackage{balance}
\usepackage{algorithm}
\usepackage{algorithmic}
\usepackage{amsmath,amssymb,amsfonts}
\usepackage{algorithmic}
\usepackage{textcomp}
\usepackage{stfloats}
\usepackage[numbers,sort&compress]{natbib}
\usepackage{balance}

\allowdisplaybreaks[4]

\newtheorem{remark}{Remark}
\newtheorem{definition}{Definition}
\newtheorem{lemma}{Lemma}
\newtheorem{proposition}{Proposition}
\newtheorem{theorem}{Theorem}
\newtheorem{corollary}{Corollary}
\newtheorem{example}{Example}

\newcommand{\norm}[1]{\left\Vert#1\right\Vert}

\newcommand{\Real}{\mathbb R}
\newcommand{\Tran}{\mathrm T}

\begin{document}

\title{Optimal Tracking Control for Unknown Linear Systems with Finite-Time Parameter Estimation}
	\author{Shengbo Wang, Shiping Wen,~\IEEEmembership{Senior Member,~IEEE}, Kaibo Shi, Song Zhu, \\ and Tingwen Huang,~\IEEEmembership{Fellow,~IEEE} 	
	
	\thanks{This publication was made possible by NPRP grant: NPRP 9-466-1-103 from Qatar National Research Fund. The statements made herein are solely the responsibility of the authors. \textit{(Corresponding authors: Shiping Wen.)}
}
	
		\thanks{S. Wang is with School of Computer Science and Engineering, University of Electronic Science and Technology of China, Chengdu 611731, China (e-mail:  shnbo.wang@foxmail.com). S. Wen is with Australian AI Institute, Faculty of Engineering and Information Technology, University of Technology Sydney, NSW 2007, Australia (shiping.wen@uts.edu.au).   K. Shi is with School of Information Science and Engineering, Chengdu University, Chengdu, 611040, China, (email: skbs111@163.com). 	S. Zhu is with School of Mathematics, China University of Mining and Technology, Xuzhou 221116, China (e-mail: songzhu@cumt.edu.cn). 
		T. Huang is with Science Program, Texas A \& M University at Qatar, Doha 23874, Qatar (e-mail: tingwen.huang@qatar.tamu.edu).}

}

	\markboth{}
{}
	\maketitle

\begin{abstract}
	The optimal control input for linear systems can be solved from algebraic Riccati equation (ARE), from which it remains questionable to get the form of the exact solution. In engineering, the acceptable numerical solutions of ARE can be found by iteration or optimization. Recently, the gradient descent based numerical solutions has been proven effective to approximate the optimal ones. This paper introduces this method to tracking problem for heterogeneous linear systems. Differently, the parameters in the dynamics of the linear systems are all assumed to be unknown, which is intractable since the gradient as well as the allowable initialization needs the prior knowledge of system dynamics. To solve this problem, the method named dynamic regressor extension and mix (DREM) is improved to estimate the parameter matrices in finite time. Besides, a discounted factor is introduced to ensure the existence of optimal solutions for heterogeneous systems. Two simulation experiments are given to illustrate the effectiveness.
\end{abstract}



\section{Introduction}\label{Section_Intro}
The research on linear controlled systems has lasted for decades and can still receive great attention nowadays. In addition to studying the internal properties such as stability and controllable, it is also hoped in some scenarios to control the system state to follow a predefined or specific trajectory, namely tracking control. Control optimization for linear systems under complex environment is an important and developing research direction \cite{optimal_linear_books}. Similarly, the optimization of tracking control is also a topic worthy of studying. The ideas for solving such problems are very clear: construct optimization goals and constraints based on the information of the systems, such as dynamics or measurable state, then obtain the optimal solution as the controller candidate. For linear systems under quadratic performance index, including the linear leader-follower systems with the goal of state tracking, the key information needed for optimization is the prior knowledge about the parameter matrices of system dynamics \cite{optimal_linear_books}. Then the optimization problem is simplified to solve algebraic Riccati equations (ARE) related to the parameter matrices \cite{optimal_linear_quadratic_books}. However, it is difficult to directly obtain the exact solution of ARE, things go even worse when the parameter matrices are unknown.

\par For unknown systems, data including measurable state, input and output can be collected to approximate some internal information of systems or to obtain the solutions of a specific target e.g. LQR performance. In terms of optimal tracking control, the error state of leader-follower systems together with designed control input can be used to analyze the effect of some candidate solutions, then choose the better solutions or adjust the solutions to perform better. For nonlinear completely unknown systems, an offline integral reinforcement learning algorithm was presented in \cite{LQRHtracking_unknownnonlinear_Jiang} to approximate the solutions with convergence and stability guaranteed. The framework of actor-critic network was used for robust tracking control with unmatched uncertainties in \cite{mu_rubust_tracking_unmatched}. Unfortunately, the implementation of this algorithm using neural networks \cite{LQRHtracking_unknownnonlinear_Jiang,mu_rubust_tracking_unmatched,mu_ADP_datadriven} is impractical because it needs to find the appropriate input features and dimensions \cite{shnbo_PETC}. Although the universal approximation ability of neural networks \cite{Universal_approximation} has been taken used in many researches on optimization for nonlinear systems \cite{LQRtracking_discounted_nonlinear_Luo,LQRtracking_two_timescale_discounted,LQR_unmeasurableD}, in research on optimization for linear cases, more practical algorithms are needed. For linear and completely unknown systems, an online model-free ADP method was introduced in \cite{linear_optimal_unknown_state_initial} to obtain the approximate solution of ARE according to the state and input. The conservatism of this method lies in full rank assumption and stabilizing initialization, which was afterwards compared with concurrent learning and interval excitation conditions in \cite{initialexcitation_comparetorank}. Same assumptions were conducted in \cite{RL_concurrent_tracking_converge} that proposed an exponential output tracking algorithm with experience replay for unknown linear systems. Besides, as an exploitation of this algorithm, the unknown parameter matrices of exosystem dynamics could be also estimated \cite{RL_concurrent_tracking_converge}. This  further exploitation inspires that: instead of directly obtaining the optimal control solution from the data, it should be better to first restore the system dynamics using data, and then approximate the optimal solutions. There are two reasons for this view. On the one hand, the data is typically the input and output of system dynamics, it is much more tractable to identify the dynamics using collective data rather than to obtain the solutions of Hamilton-Jacobi-Bellman (HJB) equation \cite{LQRHtracking_unknownnonlinear_Jiang,mu_rubust_tracking_unmatched,shnbo_PETC,mu_ADP_datadriven,track_IRL,LQR_unmeasurableD,LQRtracking_two_timescale_discounted,LQRtracking_discounted_nonlinear_Luo}. One the other hand, typically for linear systems, the more complex HJB equation is simplified to less complex ARE \cite{optimal_linear_books,optimal_linear_quadratic_books} which is decoupled from system state and input. It is much easier and optimality guaranteed to obtain the numerical solutions for linear systems. 
\par The identification for dynamics of linear systems can be transformed into the extended regression problem through filters \cite{Annual_Reviews}, estimators \cite{Rushikesh_tac_concurrent_pestimation}, etc. In order to ensure convergence, a common assumption is persistently excitation (PE) condition \cite{PEcondition}, which is unverifiable and impractical. To relax this, interval excitation was presented by \cite{IEcondition} and taken used in many researches. By mixing a data-storage based item in the updating law, namely concurrent learning, the adaptive controller as well as uncertain dynamics can be tracked with global stability under IE condition in \cite{IE_MRAC_2013}. This idea has been further presented in \cite{initialexcitation_comparetorank,Rushikesh_tac_concurrent_pestimation,RL_concurrent_tracking_converge} for approximate optimal control, while in \cite{initialexcitation_comparetorank} IE condition was compared with full rank condition \cite{linear_optimal_unknown_state_initial}. The disadvantage of concurrent learning is that, the data-storage mechanism should be intelligent enough to capture sufficiently rich data \cite{initialexcitation_comparetorank}. As another exploitation of IE condition, in the field of parameter estimation, global stability even finite-time estimation can be achieved without either full rank or data-storage assumptions. In \cite{robotic_finite_IE}, a set of filtered auxiliary variables was introduced to estimate the parameters of manipulators in finite time under IE condition. A more general technique is called dynamic regressor extension and mixing (DREM), which was proposed in \cite{DREM_begin} to estimate parameters for linear time-varying systems without PE condition. For linear time-invariant systems, DREM could improve the transient stability and realize finite-time convergence under modified IE condition \cite{matrix_estimation_finite_time}. A modified updating law of DREM with short finite-time stability was studied in \cite{finite_time_robust} in which robustness of this method was also analyzed. It is of most importance to realize the finite-time convergence for the objective of dynamics restoring and then optimal solution approximating, typically for deterministic systems. Specifically, only accurate model estimation can ensure the accuracy of numerical optimal solutions and provide validation strategy for initialization legitimacy, which is always neglected in most researches e.g. \cite{LQRHtracking_unknownnonlinear_Jiang,mu_rubust_tracking_unmatched,shnbo_PETC,LQRtracking_two_timescale_discounted,linear_optimal_unknown_state_initial,mu_ADP_datadriven}. However, DREM with finite-time convergence has not yet been introduced in general linear invariant systems except in \cite{matrix_estimation_finite_time} where DREM for multivariable systems was proposed without drift dynamics.
\par With the estimated parameters, the approximate optimal controller can be obtained by solving ARE in an offline way without additional measurement of state or input. It has been independently developed using iterative method to obtain the numerical solutions of ARE, e.g. the well known Kleinman’s algorithm \cite{KleinDARE} which needs an initial stabilizing gain. For deterministic systems with known parameters, it is verifiable and will converge to the optimal solutions. alternatively, learning methods such as policy iteration and value iteration have also been developed \cite{PI_ARE_Lee,QlearningWei}. The stability, convergence rate as well as computational complexity of these methods has been analyzed thoroughly, see \cite{DRLQLW_review} for detailed review. Conversely, the gradient based method has been rarely studied, yet it may be more suitable to approximate the solutions under constraints e.g. equality-type constraints \cite{ARE_gradient_Lsmooth}. As is already analyzed, the optimization problem with ARE can be nonconvex. Recently, a variable substitution method was proposed in \cite{ARE_gradient_variable_CDC} to reparameterize the problem into convex one, proving the global exponential convergence via gradient method. In addition, the solutions optimal output regulation using gradient method was studied in \cite{ARE_gradient_Lsmooth} according to $L-$smooth property of the objective function. What is more, the gradient can also be approximated through searching algorithm for model-free setting with convergence guaranteed \cite{model_freeLQR_TAC}. Unfortunately, these methods commonly need a stabilizing initial value, which is unverifiable for unknown systems. In terms of tracking control of heterogeneous systems, a discounted factor will be added to the original ARE, see \cite{track_IRL} for details. It is also a task to analyze applicability of solving modified ARE via gradient method for leader-follower heterogeneous systems.
\par Motivated by the observations, this paper investigates the optimal tracking control for unknown linear invariant systems using finite-time parameter estimation and gradient-based searching method. The main contributions are listed below.
\begin{enumerate}[\IEEEsetlabelwidth{12)}]
	\item The optimal tracking control for unknown heterogeneous systems is studied. The parameter estimation and numerical solutions for optimal controller have exponential convergence rate under mild conditions. In addition, the conditions on generation of signals and initialization of control gain are online verifiable, which are always neglected in most researches of unknown systems.
	\item A finite-time parameter estimation of general linear time-invariant systems with multivariables via DREM is proposed under modified interval excitation assumption. Compared with \cite{matrix_estimation_finite_time}, the drift dynamics as well as coupled terms are considered. Moreover, the modified filter-based derivative estimator is more capable than that used in \cite{matrix_estimation_finite_time,DREM_begin}, which can be used in DREM while the derivative are unmeasurable.
	\item Because of parameter mismatches of leader-follower systems, a discounted factor is introduced to ensure the stability of LQR optimization. A gradient-based method is used to approximate the optimal solutions of modified ARE with discounted factor. The criteria for exponential convergence rate and optimality is given with rigorous mathematical analysis.
\end{enumerate}
	
\par \emph{Notations}: In this paper, $\mathbb{R}$ denotes the set of real number, $\mathbb{R}^n$ and $\mathbb{R}^{m\times n}$ denote the sets of $n$ dimensional real vector and $m\times n$ dimensional real matrix, respectively. $\norm{\cdot}$ is used to denote the Frobenius norm for a vector or a matrix. For a matrix $A\in \Real^{n\times n}$,  $det\{A\}$, $tr(A)$, $adj\{A\}$, $A^\mathrm{T}$ and $A^{-1}$ represent the determinant, trace, adjoint matrix, transpose matrix and the inverse matrix of $A$, respectively. $A>0$ ($A\ge0$) means $A$ is symmetric and positive definite (semi-positive definite). $\lambda_{\max}\left(A\right)$ and $\lambda_{\min}\left(A\right)$ are the the maximum and minimum eigenvalues of $A$, respectively. $I$ and $\textbf{0}$ represent identity matrix and zero matrix with appropriate dimensions.

\section{Preliminaries}\label{Section_Pre}
In this paper, the linear time-invariant systems take the following form
\begin{align}
	\dot x(t) =& Ax(t) + Bu(t) + Cv(t), \quad x(0) = x_0\label{system_dynamics}\\
	\dot v(t) =& D w(t), \quad v(0) = v_0\label{exosystem_dynamics}
\end{align}
where $x(t)\in \Real^n$, $u(t)\in \Real^m$, $v(t)\in \Real^n$ and $w(t)\in \Real^q$ are the system state, control input, exosystem state and exosystem input, respectively. Dynamics matrix $A\in \Real^{n\times n}$ and gain matrices $B\in \Real ^{n\times m}$ $C\in \Real ^{n\times q}$ and $D\in \Real^{n\times q}$ are all unknown. For simplicity, states and inputs in autonomous systems \eqref{system_dynamics} and \eqref{exosystem_dynamics} is abbreviated as $x$, $v$, $u$ and $w$ without losing generality.
\par The goal of tracking is to regulate the state of $x$ to track the state $v$ by designing appropriate control input $u$. The rigorous definition is as follows.
\begin{definition}
	Define error state $e_s = x - v$ and the allowable error region $\Omega_e$. System $x$ is said to track system $v$ such that $\exists t_e > 0$, $\forall t>t_{e}$ there is $e_s(t) \in \Omega_e$.
\end{definition}
It might be impossible to realize $e_s(t) = 0$, $\forall t > t_e$, when the dynamics of leader signal $v$ and follower system $x$ are different and control input $u$ can not compensate for this difference. For a controllable system, an appropriate control input can narrow the error bound, i.e. $\Omega_e$. Therefore, the solution for tracking problem for heterogeneous systems is a trade-off between the power consumption of control input and error bound, which can be described by LQR problem in which the performance index takes
\begin{equation*}
	J = \int_0^\infty r(e_s,u)\, d\tau
\end{equation*}
where $r(e_s,u) = e_s^\Tran Qe_s + u^\Tran R u$, $Q\in \Real^{n\times n}$, $R\in \Real^{m\times m}$ and $Q,R>0$. For linear systems, the performance $J$ takes the quadratic form of system state or error state of leader-follower systems. It is noted that the parameter mismatches may lead $J$ to take a more complex form rather than quadratic one. The intuitive explanation is that for two systems with different drift dynamics, the control input should perform real-time compensation to force the state consensus. In this case, $J$ increases even if $\norm{e_s}\to 0$ due to $\norm{u}\ne 0$. Therefore, a modified performance index with discounted factor $\gamma > 0$ inspired by reinforcement learning has been studied in many researches \cite{track_IRL,LQRHtracking_unknownnonlinear_Jiang,LQRtracking_discounted_nonlinear_Luo,LQRtracking_two_timescale_discounted}. Then, the optimization problem is described as
\begin{equation}
	\min_{u} \quad J^\prime = \int_0^\infty e^{-\gamma} r(e_s,u)\, d\tau, \label{discounted_performance}
\end{equation}
subject to the dynamics \eqref{system_dynamics} and \eqref{exosystem_dynamics}. Then, define $J^\prime = e_s^{\Tran} P e_s$, the optimization problem can be transformed into the Hamilton-Jacobi-Bellman (HJB) equation:
\begin{equation}
	e_s^\Tran Q e_s + u^\Tran Q u + {\frac{\partial V}{\partial e_s}}^{\Tran}\dot e_s - \gamma V= 0. \label{HJB_unknown}
\end{equation}
\par It was discussed in \cite{LQRHtracking_unknownnonlinear_Jiang} that there exists an unique solution $u^*$ of \eqref{discounted_performance} as long as $\gamma$ satisfies certain bounded condition for nonlinear system. However, since the parameter matrices are all unknown, the Riccati equation that is natural simplified for linear systems is intractable to obtain. Some researches regard such unknown linear systems as affined nonlinear systems, e.g. using data-driven RL method to approximate the solution according to the HJB \eqref{HJB_unknown} as \cite{LQRHtracking_unknownnonlinear_Jiang,track_IRL}. It works but is difficult to guarantee convergence to optimal solutions. Besides, the initial control input is always assumed to be admissible, which is unverifiable under unknown circumstance. In all, these methods complicate the problem to some extend.
\par An alternative is to introduce observers to estimate the unknown parameters. At present, the research on linear regression has been more comprehensive. For a linear regression problem as
\begin{equation}
	y_r =  x_r^\Tran\theta_r, \label{pre_regressor}
\end{equation}
in which $y_r\in \Real$ and $x_r\in \Real r$ are measurable and $\theta_r\in \Real^r$ is bound but unknown, one should design a regressor $\hat \theta_r$ to estimate $\theta_r$ using collective data. Traditionally, $\hat\theta_r$ can evolve along with gradient descent direction of \eqref{pre_regressor} as
\begin{equation*}
	\dot {\hat \theta}_r = -\alpha_r x_r\left( x^{\Tran} \hat\theta_r - y_r\right) = -\alpha_r x_r x_r^\Tran \tilde \theta_r.
\end{equation*}
where $\alpha_r > 0$ represents the learning rate, $\tilde \theta_r = \hat \theta_r - \theta_r$. As $\theta_r$ is constant, one has
\begin{equation*}
	\dot {\tilde \theta}_r = -\alpha_r x_r x_r^\Tran \tilde \theta_r. \label{pre_PE_dynamics}
\end{equation*}
To ensure convergence, the measurable signal $x_r x_r^\Tran$ should be persistently excited, which is formally defined as follows.
\begin{definition}
	A signal $\phi(t)$ is said to be persistently excited (PE) if there exists $t_r>0$ and positive constant $\sigma$ such that
	\begin{equation*}
		\int_t^{t + t_r} \phi(\tau)\phi^\Tran(\tau) \, d\tau \ge \sigma I.
	\end{equation*}
\end{definition}
The PE condition is hard to achieve because it needs to activate the systems all the time which is unverifiable either. DREM relaxes this condition \cite{DREM_begin} and as can be seen in \cite{matrix_estimation_finite_time}, with IE condition defined in section \ref{Section_estimator}, finite-time parameter estimation can be achieved.
\par After estimation, HJB equation \eqref{HJB_unknown} reduces to algebraic  Riccati equation (ARE). Then, it is possible to approximate the optimal control input with guaranteed convergence rate. The most widely used method to find the numerical solution is Kleinman's algorithm \cite{KleinDARE} as follows.
\begin{lemma}
	For $\dot x = A x + B u$, let $A_k = A - B_K$, $k = 0, 1, 2\dots$. If $A_0$ is stabilizable, then with the following iteration
	\begin{align*}
		A_k^\Tran P_k + P_k A_k + Q + K_k R K_k &= 0,\\
		K_{k + 1} &= R^{-1}B^\Tran P_k,
	\end{align*}
	where $P_k$ is the  unique positive definite matrix, there is
	\begin{equation*}
		\lim_{k\to \infty} P_k = P^*,\quad \lim_{k_\infty} K_k = K^*.
	\end{equation*}
\end{lemma}
The Kleinman’s algorithm is simple and efficient with known parameter matrices in \eqref{system_dynamics}. An alternative way to approximate the optimal solutions of ARE is gradient-based method, considering $K$ as matrix variable \cite{1992_gradient_variable}. A distinct advantage of gradient method is the feasibility to solve general ARE with constraints, e.g. equality constraint \cite{ARE_gradient_Lsmooth}. In the next two sections, DREM and gradient-based optimization are developed to deal with the optimal tracking control for heterogeneous linear systems.

\section{Finite-time Parameter Estimation of System Dynamics}\label{Section_estimator}
For the unknown heterogeneous systems with measurable state and input $x$, $v$, $w$ and $u$, it can be extended to multivariable regression problem to estimate these unknown matrices. The first objective is to obtain the parameters in $D$ of drift dynamics of $v$ in finite time, dependent only on the  measurement of $v$ and $w$. Using linear filter $\mathcal{H}_0 = \frac{1}{p + \lambda}$ with $p = \frac{d}{dt}$ to obtain the regression model, one has
\begin{align}
	\dot y = & -\lambda y + \dot v, \quad y(0) = 0\\
	\dot v_l =& -\lambda v_l + v, \quad v_l(0) = 0 \\
	\dot z =& -\lambda z + w, \quad w_l(0) = 0
\end{align}
Defining $\Theta^* = D^\Tran$ induces
\begin{equation}
	y = \Theta^{*\Tran} z. \label{filtered_dynamics}
\end{equation}
Since the derivative $\dot x$ is not measurable, the filter state $y$ is solved as
\begin{equation}
	y(t) = e^{-\lambda t}\int_0^t e^{\lambda \tau} \dot v(\tau) \, d\tau,
\end{equation}
which can be further transformed into a measurable form using integration by parts as
\begin{equation}
	y(t) = v(t) - e^{-\lambda t}v(0) - \lambda v_l(t). \label{filter_derivative}
\end{equation}
\begin{remark}
	These filters are used in \cite{initialexcitation_comparetorank,RL_concurrent_tracking_converge} as the regressor of system derivative, eliminating the assumption of known derivative information \cite{DREM_begin}. Compared with the filters used in \cite{matrix_estimation_finite_time}, \eqref{filter_derivative} can track the real derivative trajectory with faster convergence speed.
\end{remark}
Introduce the one-input-$q$-output linear operators $\mathcal H[\cdot]$ as
\begin{equation}
	\mathcal{H}[\cdot] = col(\mathcal{H}_1[\cdot],\mathcal{H}_2[\cdot],\dots,\mathcal{H}_q[\cdot]),
\end{equation}
where linear operator $\mathcal{H}_i = \frac{1}{p + \lambda_i}$ is $\mathcal{L}_\infty$-stable with $\lambda_i>0$, $\lambda_i \ne \lambda_j$, $\forall i, j\in{1,\dots,q}$. Applying the operators on the transpose of \eqref{filtered_dynamics}, one has the matrix identity
\begin{equation}
	Y = \mathcal{Z} \Theta^*. \label{filtered_matrix_dynamic}
\end{equation}
with $Y = \mathcal{H}[y^\Tran]\in R^{q\times n}$ and $\mathcal{Z} = \mathcal{H}[z^\Tran] \in \Real^{q \times q}$. With the help of adjoint matrix $adj\{\mathcal{Z}\}$, define $\delta(t) = det\{\mathcal{Z}\}$ and multiply $adj\{\mathcal{Z}\}$ on the left of both sides of \eqref{filtered_matrix_dynamic} derives
\begin{equation}
	\mathcal{Y} = \delta(t) \Theta^* \label{regression_matrix}
\end{equation}
in which $\mathcal{Y} = adj\{\mathcal{Z}\} Y$. Furthermore, for $\delta$ is a scalar,
\begin{equation}
	\mathcal{Y}_{ij} = \delta(t) \Theta^*_{ij}, \quad \forall i, j\in{1,\dots,q} \label{regression_element}
\end{equation}
which is the regression model of each element in $D$. A gradient-based approach for scalar variables is feasible to approximate each parameter in $\Theta^*$. 
\begin{proposition}
	\label{proposition_1}
	Consider the linear regression function \eqref{regression_matrix} and \eqref{regression_element}. If $\delta(t)\notin \mathcal{L}_2$, i.e $\int_0^\infty \delta^2(t) \, dt = \infty$, then under the following update law,
	\begin{equation}
		\dot \Theta = \alpha \delta(t) \left(\mathcal{Y} - \delta(t) \Theta \right), \label{updatelaw_theta}
	\end{equation}
	in which $\alpha>0$ represents the learning rate, the estimation error $\tilde \Theta = \Theta - \Theta ^*$ will converge to zero. Moreover, the estimation error of each element $\tilde \Theta_{ij}$ is monotonous satisfying $\vert\tilde{\Theta}_{ij}(t_1)\vert \leq \vert\tilde{\Theta}_{ij}(t_2)\vert$, $\forall t_1 \ge t_2$.
\end{proposition}
\textbf{Proof.} Substituting \eqref{regression_matrix} into \eqref{updatelaw_theta} induces the dynamics of matrix-wise estimation error and element-wise estimation error as
\begin{equation}
	\tilde \Theta = - \alpha \delta^2(t) \tilde \Theta,\quad \tilde \Theta_{ij} = - \alpha \delta^2(t) \tilde \Theta_{ij}. \label{thetaerror_dynamics}
\end{equation}
Solving the differential equations yields
\begin{equation}
	\tilde{\Theta}_{ij}(t) = e^{-\alpha \int_0^t \delta(\tau)^2\, d\tau} \tilde{\Theta}_{ij}(0). \label{convergence_of_each_element}
\end{equation}
As $\delta^2(t) \ge 0$, $\tilde{\Theta}_{ij}(t)$ is monotonic non-increasing. When $\delta(t)\notin \mathcal{L}_2$, all the elements in $\tilde{\Theta}$ converge to zero, i.e. $\tilde{\Theta}$ will converge to zero.
\begin{remark}
	When discussing the conservatism of PE, it should be noted that PE condition ensures strong regression ability, specifically, exponential stability of estimating the time-varying parameters. As proven in \cite{DREM_begin}, it is of asymptotical stability for DREM to estimate time-varying parameters with $\delta^2(t) \notin \mathcal{L}_2$. In addition, it was discussed that $\delta^2(t) \notin \mathcal{L}$ is weaker that $z(t)$ satisfying PE condition. It can be verified by considering $\delta(t) = \frac{1}{\sqrt{t + 1}}$, evidently $\delta^2(t)\notin \mathcal{L}_2$ but not satisfies PE condition.
\end{remark}
For linear time-invariant systems, the PE condition as well as $\delta^2(t) \notin \mathcal{L}_2$ is over conservative to estimate the unknown parameters. It is shown next that the parameter matrices in drift dynamics of multivariable systems can be estimated in finite time interval as long as $\delta(t)$ satisfies a modified IE condition which is formally defined below.
\begin{definition}
	A signal $\phi(t)$ is said to be interval excitation (IE) if there exists time instance $t_c>0$ and positive constant $\sigma$ such that
	\begin{equation}
		\int_0^{t_c} \phi(\tau)\phi^\Tran(\tau) \, d\tau \ge \sigma I.
	\end{equation}
	To further introduce the finite-time convergence of the parameter estimation, a modified IE condition can be defined as
	\begin{equation}
		\int_0^{t_c} \phi(\tau)\phi^\Tran(\tau) \, d\tau \ge -\frac{1}{\alpha}\ln\left(1-\sigma\right), \label{IEcondition_ln}
	\end{equation}
	where $\sigma\in(0,~1)$ and $\alpha$ is the learning rate defined in \eqref{updatelaw_theta}.
\end{definition}

\begin{proposition}\cite{matrix_estimation_finite_time}
	\label{prop2}
	Define the scalar signal $s_0$ with the following dynamics
	\begin{equation}
		\dot s_0(t) = -\alpha \delta^2(t) s_0(t),\quad s_0(0) = 1. \label{aux_signal}
	\end{equation}
	Consider the linear regression function \eqref{regression_matrix} and \eqref{regression_element}. With the updating law \eqref{updatelaw_theta}, if $\delta(t)$ satisfies IE condition \eqref{IEcondition_ln}, the elaborate parameter matrix $\Theta_F$ with the following definition
	\begin{equation}
		\Theta_F = \frac{1}{1-s}\left(\Theta - s\Theta(0) \right) \label{finitetime_law}
	\end{equation}
	will converge to $\Theta^*$ within $(0,t_c)$, where $s$ is the continuous signal defined by
	\begin{align}
		s = \left\{\begin{array}{lll}
				1 - \sigma, & if & s_0(t) > 1-\sigma,\\
				s_0(t), & if & s_0(t) \leq 1 - \sigma.
			\end{array}\right. \label{switching_condition}
	\end{align}
\end{proposition}
\textbf{Proof.} It is noticed that $s\leq 1 - \sigma< 1$. When $t\in [0, t_c)$, one has
\begin{align*}
	\Theta_F = &\frac{1}{\sigma}\left(\Theta - (1 - \sigma)\Theta(0) \right),
\end{align*}
which implies no finite escape time in the calculation process of $\Theta_F$. Solving the differential equation \eqref{aux_signal} yields
\begin{equation}
	s_0(t) = e^{-\alpha \int_0^t \delta(\tau)^2\, d\tau} s_0(0) = e^{-\alpha \int_0^t \delta(\tau)^2\, d\tau}.
\end{equation}
According to \eqref{convergence_of_each_element}, there is
\begin{equation}
	\tilde{\Theta} = s_0(t) \tilde{\Theta} ~\Longrightarrow~ \Theta - \Theta^* = s_0(t) \left(\Theta(0) - \Theta^* \right).
\end{equation}
When IE condition \eqref{IEcondition_ln} is met and $t\ge t_c$, i.e. $s_0 \leq 1 - \sigma$, one has $s = s_0(t)$. The original parameter matrix can be obtained by
\begin{equation}
	\Theta^* = \frac{1}{1 - s} \left( \Theta - s\Theta(0) \right) = \Theta_F,
\end{equation}
which implies that $\Theta_F$ converges to $\Theta^*$, $\forall t \ge t_c$.
\begin{lemma}
	There exist some other methods to realize finite-time parameter estimation based on DREM as studied in \cite{finite_time_robust}. The idea is simple but effective: modify the updating law of estimator \eqref{updatelaw_theta} into an adaptive law with low order feedback, like
	\begin{equation*}
		\dot \Theta = \alpha \delta(t) \left(\mathcal{Y} - \delta(t) \Theta \right)^o,\quad o \in (0,1]
	\end{equation*}
	such that the tracking effect is strengthened when the tracking error becomes small, e.g. when $\tilde{\Theta}_{ij}\in (-1,1)$. Therefore, finite-time convergence is ensured solving modified \eqref{thetaerror_dynamics}. The robustness of this method was analyzed in \cite{finite_time_robust} under disturbances. However, these methods are more complicated than that proposed in Proposition \ref{prop2}, which is more difficult to be introduced to general heterogeneous linear systems \eqref{system_dynamics}. 
\end{lemma}
So far, the parameter matrix in drift dynamics of linear systems can be estimated as illustrated. It is also feasible to estimate the coefficient matrices $A$, $B$ and $C$ using the same method. To obtain the information of derivative $\dot x(t)$, the filters are applied as
\begin{align}
	\dot \psi_x =& -\lambda \psi_x + x, \quad \psi_x(0) = 0, \\
	\dot \psi_u =& -\lambda \psi_u + u, \quad \psi_u(0) = 0, \\
	\dot \psi_v =& -\lambda \psi_v + v, \quad \psi_v(0) = 0.
\end{align}
Then, defining $\psi_y = -\lambda \psi_x + x$, one has
\begin{equation}
	\psi_y = A \psi_x + B\psi_u + C\psi_v = \Psi^{*\Tran} \psi_z, \label{linear_regression}
\end{equation}
where $\Psi^* = [A^\Tran,~B^\Tran,~C^\Tran]^\Tran$, $\psi_z = [\psi_x^\Tran, \psi_u^\Tran, \psi_v^\Tran]^\Tran$. Introducing the one-input-$(n+m+q)$-output linear operators $\mathcal H_\psi[\cdot]$ and following the same transformation, one has the matrix-wise regression function
\begin{equation}
	\mathcal{E} = \Delta(t) \Psi^*,
\end{equation}
with $\mathcal{E}= adj\left\{F \right\} \mathcal{H}_\psi[\psi_y^\Tran]\in R^{(n+m+q)\times n}$, $F = \mathcal{H}_\psi[\psi_z^\Tran] \in \Real^{(n+m+q) \times (n+m+q)}$ and $\Delta(t) = det\left\{F \right\}$, and the element-wise regression function
\begin{equation}
	\mathcal{E}_{ij} = \Delta(t) \Psi^*_{ij},\quad \forall i, j\in{1,\dots,(n+m+q)}.
\end{equation}
\begin{theorem}
	For linear deterministic systems \eqref{system_dynamics}, the unknown parameter matrices $A$, $B$ and $C$, alternatively $\Psi^*$, can be estimated asymptotically if $\Delta \notin \mathcal{L}_2$ with the following law
	\begin{equation}
		\dot \Psi = \alpha \Delta(t) \left(\mathcal{\mathcal{E}} - \Delta(t) \Psi \right), \label{linear_regression_updatelaw}
	\end{equation}
	where $\Psi$ as the estimation of the coefficient matrix $\Psi^*$. Furthermore, the finite-time convergence can be guaranteed if $\Delta(t)$ satisfies IE condition \eqref{IEcondition_ln} with \eqref{aux_signal}, \eqref{switching_condition} and
	\begin{equation}
		\Psi_F = \frac{1}{1-s}\left(\Psi - s\Psi(0) \right), \label{intergrated_finiteestimation}
	\end{equation}
	where $\Psi_F$ is the finite-time estimated matrix of $\Psi^*$.
\end{theorem}
The proof process is the same as that in Proposition \ref{proposition_1} and \ref{prop2}, which is omitted for brevity.
\begin{remark}
	The finite-time estimation \eqref{intergrated_finiteestimation} can be computed only once when $t = t_c$, which means that $\Delta(t)$ satisfies modified IE condition \eqref{IEcondition_ln} within $t\in (0,t_c]$. The reverse calculation of \eqref{intergrated_finiteestimation} can be more interesting. Suppose that $\sigma$ in modified IE condition \eqref{IEcondition_ln} is not available at the beginning of estimation. Finite-time estimation in $(0,t_c]$ can be still achieved if $\Delta(t)$ is excited. When $t = t_c$, one can calculate $\sigma = 1 - s_0(t)$ inversely, and assign $s_0$ and $\Psi_F$ accordingly, which is a kind of potential high-gain injection. This holds true only if the signal $\Delta(t)$ satisfies IE condition and $t_c$ is set not too small, or there exists finite escape time. As $s_0$ is computed online, the IE condition is online verifiable.
\end{remark}

\section{Finite-time Parameter approximation of Optimal Controller}\label{Section_opt}
Since the derivative of error state $e_s$ contains the state $v$, which is not feasible to obtain ARE directly. Therefore, the augmented systems are given by
\begin{align}
	\dot X = \mathcal{A} X + \mathcal{B} U, \quad X_0 = col\left(x_0 - v_0,v_0\right) \label{augmented_system_dynamics}
\end{align}
where 
\begin{align*}
	X = col\left(e_s,v\right), ~
	\mathcal{A} = \left[\begin{matrix}
		A & A+C-D\\
		\textbf{0} & D
	\end{matrix}\right],~\mathcal{B} = \left[\begin{matrix}
		B\\
		\textbf{0}
	\end{matrix}\right]
\end{align*}
and $U = u$. The tracking value function is defined as
\begin{equation}
	J\left(X,U \right) = \int_0^\infty e^{-\gamma\tau} r^\prime(X(\tau),U(\tau))\, d\tau \label{LQR_performance}
\end{equation}
where $ r^\prime(X,U) = X^\Tran \hat Q X + U^\Tran R U$ with $\hat Q = diag\{Q,\textbf{0}\}$ and $\gamma>0$ is the discounted factor for stability. For linear systems with linear feedback controller $U^* = - K^* X$, the value function takes the quadratic form as
\begin{equation}
	J\left(X,U \right) = f(K) = X_0^\Tran P(K) X_0 
\end{equation}
where $f(K)$ is the function of $K$ to be minimize and $P(K)>0$ is the solution of the following algebraic Riccati equation (ARE)
\begin{align}
	P(K) \left(\mathcal{A} - \mathcal{B}K\right)& + \left(\mathcal{A} - \mathcal{B}K\right)^\Tran P(K)\nonumber\\
	& - \gamma P(K) + K^\Tran RK + \hat Q = 0, \label{ARE_TRACKING}
\end{align}
where $K$ should be in the set of stabilizing gains defined as
\begin{equation}
	\mathcal{S} = \left\{ K\in \Real^{m \times 2n} ~\vert ~ \left(\mathcal{A} - \mathcal{B}K - 0.5 \gamma I\right) ~ \text{is Hurwitz}\right\}.
\end{equation}
According to the stationarity condition of optimality, the optimal controller is
\begin{equation}
	K^* = R^{-1} \mathcal{B}^\Tran P^*.
\end{equation}
Then, the ARE equation is transformed into
\begin{equation}
	P^* \mathcal{A} + \mathcal{A}^\Tran P^*
	 - \gamma P^* - P^*\mathcal{B}R^{-1}\mathcal{B}^\Tran P^* + \hat Q = 0. \label{ARE_P}
\end{equation}
It is discussed in \cite{track_IRL} that ARE \eqref{ARE_P} has a unique solution if $\left(\mathcal{A}-0.5\gamma I, \mathcal{B}\right)$ is stabilizable.
However, it is difficult to get the explicit form of $P^*$. The gradient-based optimization method encounters the fact that the feasible solution domain $\mathcal{S}$ can be nonconvex with non-smooth boundary \cite{ARE_gradient_Lsmooth}. 
\par It is proven in \cite{ARE_gradient_variable_CDC,1992_gradient_variable} that the LQR problem of linear systems can have convex representation through variable changes within an admissible domain. With the help of finite-time parameter estimation, it is possible to study the specific search space instead of entire parameter space to ensure the convergence and optimality of the gradient method. Specifically, the LQR problem can be solved using direct search for the optimal control gain matrix, as
\begin{align}
	\min_K \quad f(K)
\end{align}
where
\begin{align*}
	f(K) = \left\{\begin{matrix}
			tr\left( P(K)  \Pi\right),  & K\in \mathcal{S} \\
			+ \infty, & \text{otherwise,}
		   \end{matrix}\right.
\end{align*}
$\Pi = X_0 X_0^\Tran$ and $P(K) = \int_0^\infty e^{\mathscr{A}^\Tran\tau} \Omega e^{\mathscr{A}\tau} \, d\tau$ is the unique positive definite solution of Lyapunov equation \eqref{ARE_TRACKING} with $\mathscr{A} = \mathcal{A} - \mathcal{B}K - 0.5 \gamma I,~ \Omega =  K^\Tran RK + Q$. To optimize this problem,  the gradient of $f(K)$ to $K$ is computed as
\begin{equation}
	\nabla_{K} f(K) = 2 (RK - \mathcal{B}^\Tran P) Z \label{gradientoffK}
\end{equation}
where operator $\nabla_{K} = d/d K$ and $Z$ is the solution of $2(\mathcal{A} - \mathcal{B}K)Z=-\Pi$ \cite{ARE_gradient_Lsmooth}. Performing gradient descent on $K$ obtains the continuous-time searching policy
\begin{equation}
	\dot K = -\nabla_K f(K),\quad K(t_0) = K_0, \label{search_alg}
\end{equation}
where $t_0\ge 0$ is the beginning time point of optimization which is unnecessary to be $0$. $K_0\in \Real^{m\times n}$ is the initial control gain.
\par As stated above, the search space $\mathcal{S}$ is nonconvex and descent method e.g. gradient descent may not find the optimal solution. With $\mathcal{A}$, $\mathcal{B}$ estimated within finite time interval $(0,t_c)$, it is feasible to set the initial control gain $K_0\in \mathcal{S}$ and $t_0 = t_c$. On the contrary, initial value $K_0$ is unverifiable in $\mathcal{S}$ for unknown systems even if the parameters is approximated with exponential rate, owning to $t_0 \to +\infty$. A new search space $\mathcal{S}_{K_0}$ can be set with the help of initial stabilizing control gain $K_0$ as
\begin{equation}
	\mathcal{S}_{K_0} = \ \left\{ K\in \mathcal{S} ~\vert ~ f(K)\leq f(K_0) \right\}.
\end{equation}
\par The following statement referred from \cite{convergence_lemma} will be used to analyze the convergence and optimality of \eqref{search_alg}.
\begin{lemma}\cite{convergence_lemma}
	\label{lemma_optimal}
	For a continuous, twice differentiable functional $f(x(t))$ defined in Hilbert space whose derivative is defined by $g(x(t))$, if there exists a minimum value $f^* = \inf f(x^*)\ne -\infty$, the gradient descent iterative policy $\dot x = -g(x(t))$ will lead to
	\begin{equation}
		\lim_{t\to \infty} g(x(t)) = 0, \quad \norm{x(t) - x^*}\leq \iota\rho e^{-rt}
	\end{equation}
	in the region $\mathcal{S}^{\prime} = \left\{ x \vert \norm{x(t) - x(0)} \leq \rho \right\}$  provided that the following conditions are met
	\begin{align}
		\norm{g(x(t_1)) - g(x(t_2))}\leq& R \norm{x(t_1) - x(t_2)},\label{condLp_lemmma_opt}\\
		\norm{g(x(t))}^2 \ge& 2 r \left(f(x(t)) - f^*\right),\label{condLPL_lemmma_opt}
	\end{align}
	where $r,R,\rho > 0$ and $\iota = \frac{\sqrt{2R \left(f(x(0)) - f^*\right)}}{r\rho}\leq 1$. It means that $g(x(t))$ is Lipschitz-continuous \eqref{condLp_lemmma_opt} and satisfies Polyak-Lojasiewicz (PL) condition \eqref{condLPL_lemmma_opt}.
\end{lemma}
It is shown in \cite{model_freeLQR_TAC} that $\mathcal{S}$ is connected. Moreover, the new search space $\mathcal{S}_{K_0}$ is in $\mathcal{S}$ and the constraint $tr(P(K)\Pi)\leq tr(P(K_0)\Pi)$ is convex such that the new set $\mathcal{S}_{K_0}$ is connected as well. It is also shown in \cite{ARE_gradient_Lsmooth} that $\mathcal{S}_{K_0}$ is bounded and strongly convex in the neighborhood of $K^*$. Because of the connectivity and boundedness of $\mathcal{S}_{K_0}$ and $\mathcal{S}$, $x^*$ is in $\mathcal{S}_{K_0}$ and should not be in $\mathcal{S}\setminus\mathcal{S}_{K_0}$. Therefore, the defined search space is sufficiently satisfying the requirement in Lemma \ref{lemma_optimal}.
\begin{proposition}\cite{ARE_gradient_Lsmooth}
	\label{proposition_LPL}
	Assume that $\Pi$ is positive definite. For $K_0\in \mathcal{S}$ and $K\in \mathcal{S}_{K_0}$, the function $\nabla_Kf(K)$ is $L-$smooth (namely $\nabla_K f(K)$ is Lipschitz-continuous) with Lipschitz constant
	\begin{equation}
		R = \frac{2f(K_0)}{\lambda_{\min}(Q)} \left(\lambda_{\max} (R) + \norm{\mathcal{B}} \zeta \right)
	\end{equation}
	and satisfies
	\begin{align}
		& \frac{\left(\norm{\mathcal{A}- 0.5\gamma I} + \norm{K}\norm{\mathcal{B}}\right)^2\lambda_{\max}(S)}{\lambda_{\min}(R)\lambda_{\min}(\Pi)}\norm{\nabla_{K}f(K)}^2\nonumber\\
		\ge & f(K) - f(K^*) \label{LPL_condition_fK}
	\end{align}
	where $\zeta = \frac{\sqrt{n}f(K_0)}{\lambda_{min}(\Pi)}\left(\Gamma_{K_0}+ \sqrt{\left( \Gamma_{K_0}^2 + \lambda_{\max}(R) \right)^2} \right)$, $\Gamma_{K_0} = \frac{f(K_0)\norm{\mathcal{B}}}{\lambda_{\min}(\Pi)\lambda_{\min}(Q)} $ and $S>0$ is the solution of
	\begin{equation}
		\left(\mathcal{A}-\mathcal{B}K \right) S + S \left(\mathcal{A}-\mathcal{B}K \right)^\Tran - \gamma S + \Pi = 0. \label{LPcondition1}
	\end{equation}
\end{proposition}
\textbf{Proof.} The main analysis process can be found in \cite{ARE_gradient_Lsmooth}. One difference is that a discounted factor $\gamma$ is introduced in the LQR modeling \eqref{LQR_performance} for the augmented systems \eqref{augmented_system_dynamics}. Following the same process, one can denote $A_\gamma = \mathcal{A} - \frac{1}{2}\gamma I$ and obtain the same results by replacing $A$ with $A_\gamma$. Another difference is that this paper only considers the Frobenius norm due to simplicity which can be extended to obtain a tighter bound if needed.
\par The PL condition in Lemma \ref{lemma_optimal} is satisfied as discussed in the following statement by finding the upper bound of $\lambda_{\max}(S)$ and $\norm{K}$.
\begin{corollary}
	Assume that $\Pi$ is positive definite. For $K_0\in \mathcal{S}$ and $K\in \mathcal{S}_{K_0}$, the function $f(x(t))$ satisfies PL condition as
	\begin{equation}
		\norm{\nabla_K f(K)}^2 \ge 2 \mu \left(f(K) - f(K^*)\right), \label{LPL_CONDITIONS}
	\end{equation}
	where 
	\begin{equation*}
		\mu = \frac{\lambda_{min}(R)\lambda^2_{\min}(\Pi)\lambda_{\min}(Q)}{{8f(K_0)\left(\norm{\mathcal{A}} + 0.5\gamma n + \frac{\norm{\mathcal{B}}^2f(K_0)}{\lambda_{\min}(\Pi) \lambda_{\min}(R)}\right)}}.
	\end{equation*}
\end{corollary}
\textbf{Proof.} According to \eqref{LPcondition1}, $f(x(t))$ can be written as
\begin{equation}
	f(K) = tr(P(K)\Pi) = tr\left(S\left(Q + K^\Tran R K \right)\right).
\end{equation} 
For positive semi-definite matrices $X$ and $Y$, there is 
\begin{equation*}
	\lambda_{\min}(X)tr(Y)\leq tr(XY)\leq \lambda_{\max}(X)tr(Y).
\end{equation*} 
Therefore, on the one hand,
\begin{align*}
	f(K) 
\ge& \lambda_{\min}\left( Q+K^\Tran RK \right) tr(S)\\
 \ge& \lambda_{\min}\left( Q+K^\Tran RK \right)\lambda_{\max}\left( S\right),
\end{align*} 
which yields
\begin{align}
	\lambda_{\max}\left( S\right)\leq \frac{f(K)}{\lambda_{\min}\left( Q+K^\Tran RK \right)}. \label{scale_1}
\end{align} 
On the other hand, 
\begin{align*}
	f(K) 
\ge& tr\left( S K^\Tran R K \right) \ge \lambda_{\min}\left( S \right) \lambda_{\min}\left( R \right) \norm{K}^2. 
\end{align*}
According to \eqref{LPcondition1} and Lyapunov equation \cite{Lyapunov_equation}, 
\begin{align*}
	\lambda_{\min}(S) \ge &\frac{\lambda_{\min}\left( \Pi \right)}{2\norm{\mathcal{A} - 0.5\gamma I - \mathcal{B}K}}\\
	\ge & \frac{2\lambda_{\min}\left( \Pi \right)}{4\norm{\mathcal{A}} + \gamma n + 2\norm{\mathcal{B}}\norm{K}}.
\end{align*}
Then, a scalar quadratic inequality of $\norm{K}$ is obtained as
\begin{align*}
	\norm{K}^2 - b \norm{K} - c\leq 0
\end{align*}
where $b = {2\norm{\mathcal{B}}f(K)}/\left({\lambda_{\min}\left( \Pi \right)\lambda_{\min}\left( R \right) }\right)$ and $c = {2\left(\norm{\mathcal{A}} + 0.5\gamma n\right)f(K)}/\left({\lambda_{\min}\left( \Pi \right)\lambda_{\min}\left( R \right)}\right)$. Solving this inequality to obtain the largest solution of $\norm{K}$, one has
\begin{align}
	\norm{K} \leq & \frac{b + b\sqrt{1 + 4 c/b^2}}{2} \nonumber\\
	\leq & \frac{2\norm{B} f(K)}{\lambda_{\min}(\Pi)\lambda_{\min}(R)} + \frac{2\norm{\mathcal{A}} + \gamma n}{2\norm{B}}, \label{scale_2}
\end{align}
taking the fact that $\sqrt{1 + a}\leq 1 + a$ if $a \ge 0$. Notice that $f(K_0)\ge f(K)$, $\forall K \in \mathcal{S}_{K_0}$. Substituting \eqref{scale_1} and \eqref{scale_2} into \eqref{LPL_condition_fK} can result in the PL condition \eqref{LPL_CONDITIONS}. The proof is complete.
\par It has been shown that in the search space $\mathcal{S}_{K_0}$, the objective function $f(K)$ is $L-$smooth and satisfies PL inequality \eqref{condLPL_lemmma_opt}. The convergence and optimality of the gradient descent search algorithm \eqref{search_alg} is ensured in the following theorem.
\begin{theorem}
	\label{theorem2}
	Assume that $\Pi$ is positive definite. For $K_0\in \mathcal{S}$ and $K\in \mathcal{S}_{K_0}$, there is $\lim_{t\to \infty} \nabla_Kf(K) = 0$ and
	\begin{equation}
		\norm{K - K^*} \leq \iota \rho e^{-R(t-t_0)},\quad {\phi(t)\leq \phi(0) e^{-2R(t- t_0)}}
	\end{equation}
	where $\iota = \frac{\sqrt{2R\phi(t_0)}}{\mu \rho}$ and $\phi(t) = f(K(t)) - f(K^*)$.
\end{theorem}
\textbf{Proof.} (Convergence) As the descent algorithm is based on gradient, taking the derivative of $f(K(t))$, one has
\begin{equation}
	\frac{d}{dt} f(K(t)) = {\frac{d}{dt}} K^\Tran \nabla_K f(K)  = -\norm{\nabla_K f(K)}^2 \leq 0,
\end{equation}
which implies that $f(K)$ is monotonic non-increasing and $K\in \mathcal{S}_{K_0}$. Define $\underline{K} \in \mathcal{S}_{K_0}$ and ${f(\underline{K})}\ge f(K^*)$ be the minimum value of $f(K)$ while searching. Within any neighborhood $\mathcal{B}(\underline{K})=\{ K\vert K \in (\underline{K}+\Delta),\norm{\Delta}\leq \epsilon,\epsilon>0\}$, for $\Delta K \in \mathcal{B}_{\underline{K}}$ there is
\begin{equation*}
	f(\underline{K}+\Delta K) = f(\underline{K}) + \Delta^\Tran K \nabla_K f(\underline{K}).
\end{equation*}
Selecting the direction $\Delta K = -\nabla_K f(\underline{K})$ results in 
\begin{equation*}
	f(\underline{K}+\Delta K) = f(\underline{K}) -\norm{\nabla_K f(\underline{K})}^2\leq f(\underline{K}).
\end{equation*}
It is obvious that $\norm{\nabla_K f(\underline{K})} = 0$. Since $f(K)$ is twice differentiable, $K\to \underline{K}\Rightarrow \nabla_Kf(K) \to 0$. 
\par (Optimality) Denote $\phi(t) = f(K(t)) - f(K^*)$ whose derivative is given by
\begin{equation*}
	\frac{d}{dt}\phi(t) = -\norm{\nabla_Kf(K)}^2\leq -2\mu \phi(t).
\end{equation*}
Using comparison principle gives $\frac{d}{dt}\phi(t)  \leq \phi(t_0) e^{-2\mu (t-t_0)}$ reflecting that $f(K)$ converges to $f(K^*)$. Meanwhile, for $t_1>t_2 \ge t_0$, one has
\begin{align*}
	\norm{K(t_1) - K{(t_2)}} \leq & \int_{t_1}^{t_2}\norm{\nabla_Kf(K(\tau))}\, d\tau\\
	\leq & \sqrt{2R} \int_{t_1}^{t_2}\sqrt{\phi(t)}\, d\tau\\
	\leq& \frac{2R\phi(t_0)}{\mu}\left( e^{-\mu (t_1-t_0)} - e^{-\mu (t_2-t_0)}\right),
\end{align*}
which implies that $\norm{K(t)-K^*}\leq \frac{\sqrt{2R\phi(0)}}{\mu} e^{-\mu (t-t_0)}$ since there is only one optimal solution of $K^*$ when PL inequality is satisfied.
\par The optimal tracking control protocol and learning algorithm are listed in Algorithm \ref{alg_integrated}. 
\begin{remark}
	Since $f(K)$ is strongly convex in the neighborhood of $K^*$, it is suggested to improve the gradient-based updating law \eqref{gradientoffK} by other efficient or robust gradient-based methods such as adaptive gradient methods \cite{adgrad}.
\end{remark}

\section{Simulations}

\begin{algorithm}[t]
	\caption{Optimal Tracking Control Algorithm for Unknown Linear Systems}
	\textbf{Input}: dimensions of $A$, $B$, $C$ and $D$ are given, states of $x$, $v$ and $\omega$ are measurable and $u$ is designed to satisfy IE condition.
	\textbf{Output}: Optimal Tracking Controller $u^*$.

	\emph{Step 1. Finite-time Parameter estimation.}
	
	\par $\quad$ Set $\psi \in \Real^{k \times n}$ where $ k = 2n + m + q$ with initial value $\psi(0) = \psi_0$. Introduce one-input-$k$-output linear operator $\mathcal{H}_{\psi}$ and get $F = \mathcal{H}_{\psi}\left[ \psi \right] \in \Real^{k\times k}$.
	\par $\quad$ \textbf{while} $t\in [0,t_c]$:
	\par $\quad$ $\quad$ Calculate adjoint matrix $adj \left\{F \right\}$ and $\Delta = det\left\{F \right\}$.
	\par $\quad$ $\quad$ Determine the learning rate $\alpha>0$ for descent updating policy \eqref{linear_regression_updatelaw} and record the signal state $s$ according to \eqref{switching_condition}.
	\par $\quad$ $\quad$ \textbf{if} $t == t_c$
	\par $\quad$ $\quad$ $\quad$ Determine the parameter matrices by $\psi_F$ with \eqref{intergrated_finiteestimation}.
	\par $\quad$ $\quad$ \textbf{end if}
	\par $\quad$ \textbf{end while}

	\emph{Step 2. Optimal Gain Matrix Searching.}
	\par $\quad$ Initialize $K(t_c) = K_0\in \mathcal{S}_0$ with estimated parameter matrices.
	\par $\quad$ Decide the allowable error bound $\epsilon_b$
	\par $\quad$ \textbf{while} $\norm{\nabla_K f(K)} > \epsilon_b$
	\par $\quad$ $\quad$ Calculate $P(K)$ and $Z$ according to \eqref{ARE_TRACKING} and \eqref{gradientoffK}.
	\par $\quad$ $\quad$ Update $K$ with gradient descent law \eqref{search_alg}
	\par $\quad$ \textbf{end while}
	
	\emph{Step 3. Optimal Controller Design.}
	\par $\quad$ Record the value of $K$.
	\par $\quad$ The optimal controller is $u^* = -Kcol\left(x - v,v\right)$.
	\label{alg_integrated}
\end{algorithm}

\begin{figure}[t]
	\centering
	\subfigure[]{
		\begin{minipage}{4.1cm}
		\centering
		\includegraphics[width=4.1cm]{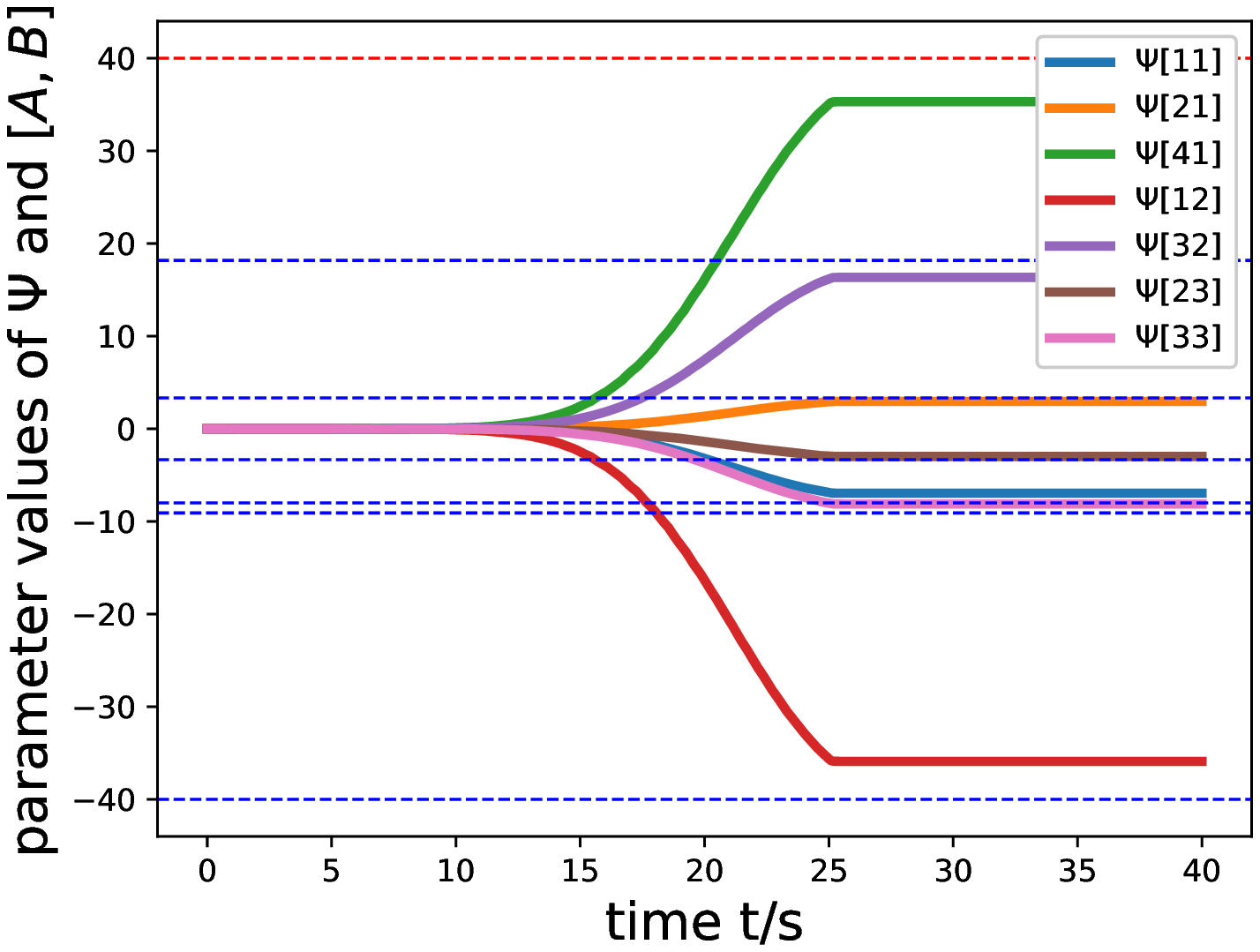}
		\end{minipage}
		\label{ex1_fig_delta_1}
		}
	\subfigure[]{
		\begin{minipage}{4.1cm}
		\centering
		\includegraphics[width=4.1cm]{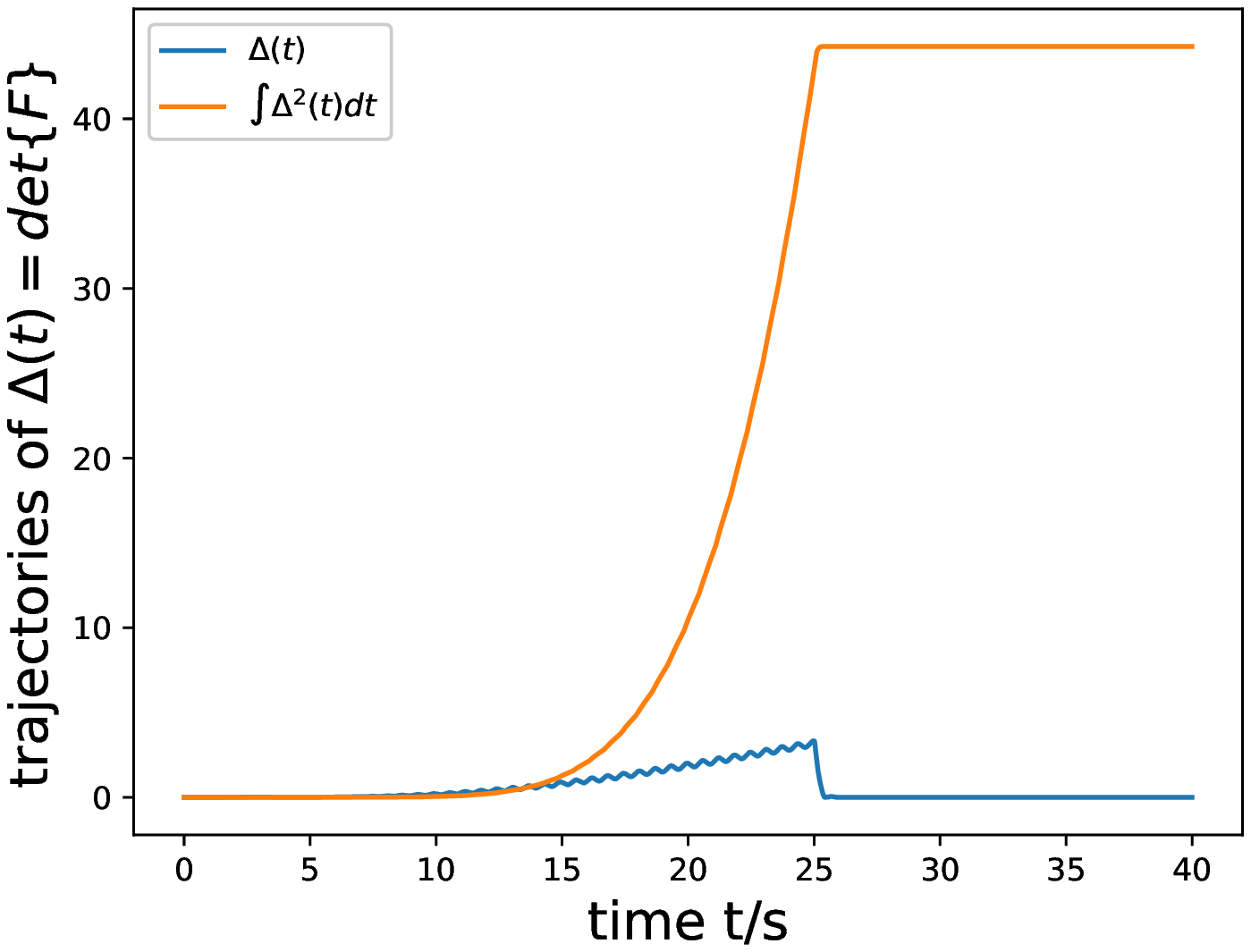}
		\end{minipage}
		\label{ex1_fig_delta_2}
		}
	\subfigure[]{
		\begin{minipage}{4.1cm}
		\centering
		\includegraphics[width=4.1cm]{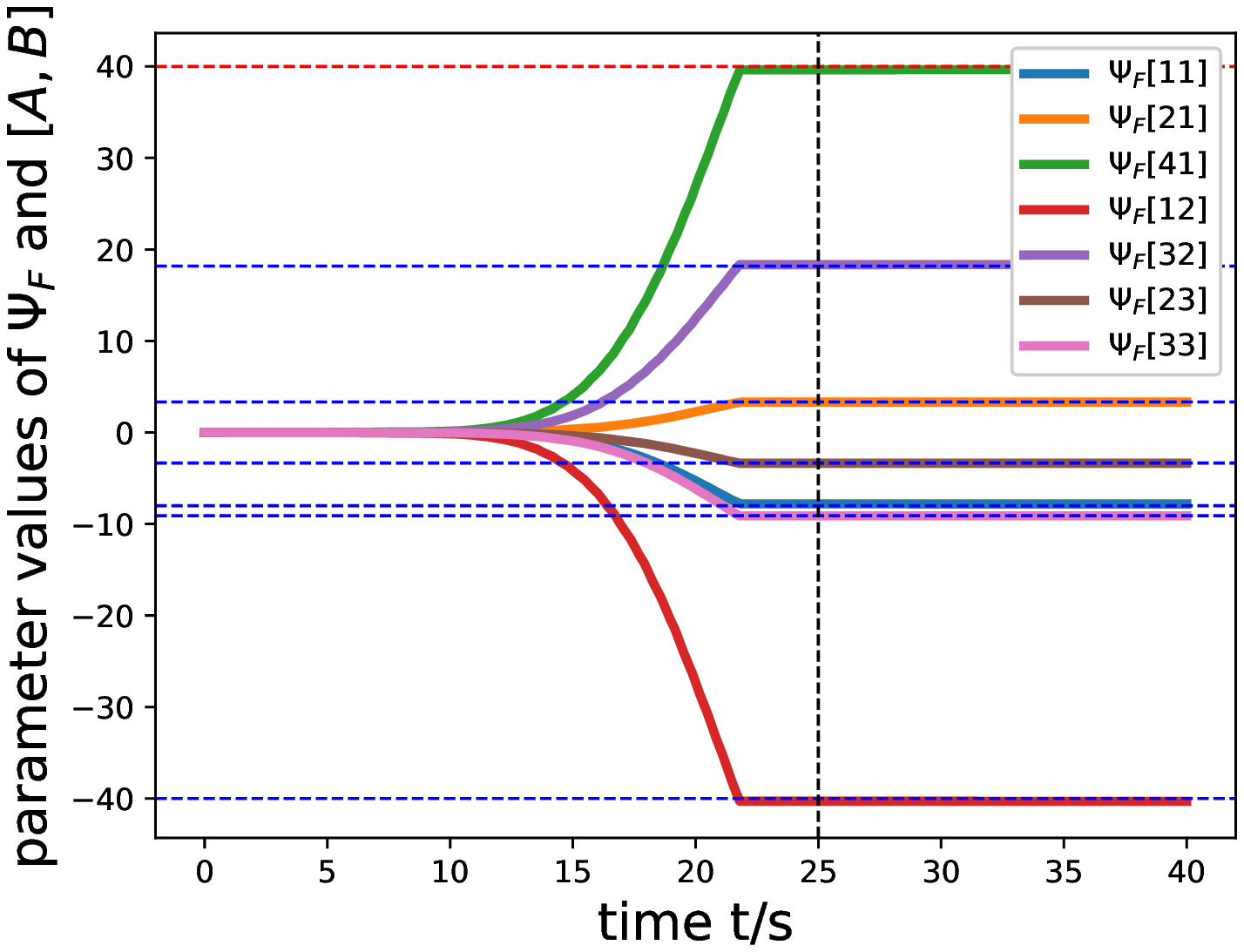}
		\end{minipage}
		\label{ex1_fig_delta_3}
		}
	\subfigure[]{
		\begin{minipage}{4.1cm}
		\centering
		\includegraphics[width=4.1cm]{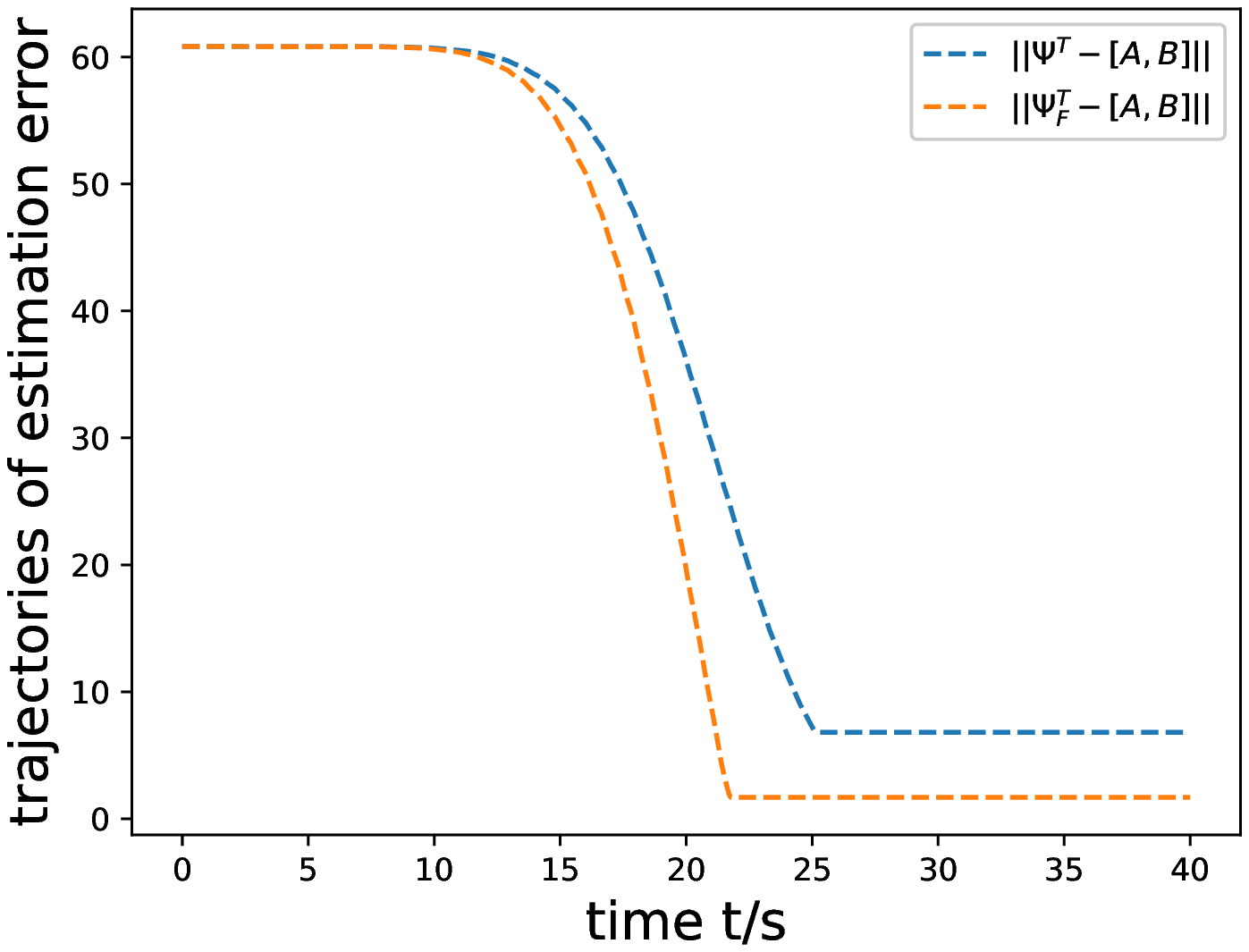}
		\end{minipage}
		\label{ex1_fig_delta_4}
		}
		\caption{Signals' and estimation parameters' trajectories in Example \ref{ex1}. (a) Parameter estimation using updating law \eqref{linear_regression_updatelaw}, where the blue dotted lines are the nonzero values of parameters in $A$ and the red dotted line represents nonzero value in $B$.  (b)  $\Delta(t)\in\mathcal{L}_2$ satisfying IE condition. (c) Parameter convergence under \eqref{finitetime_law} within $40s$, where the black line represents the maximum allowable time point. (d) Parameter estimation errors under updating law \eqref{linear_regression_updatelaw} and finite-time law \eqref{finitetime_law}.}
		\label{ex1_fig_delta}
\end{figure}

\begin{figure}[t]
	\centering
	\subfigure[]{
		\begin{minipage}{4.1cm}
		\centering
		\includegraphics[width=4.1cm]{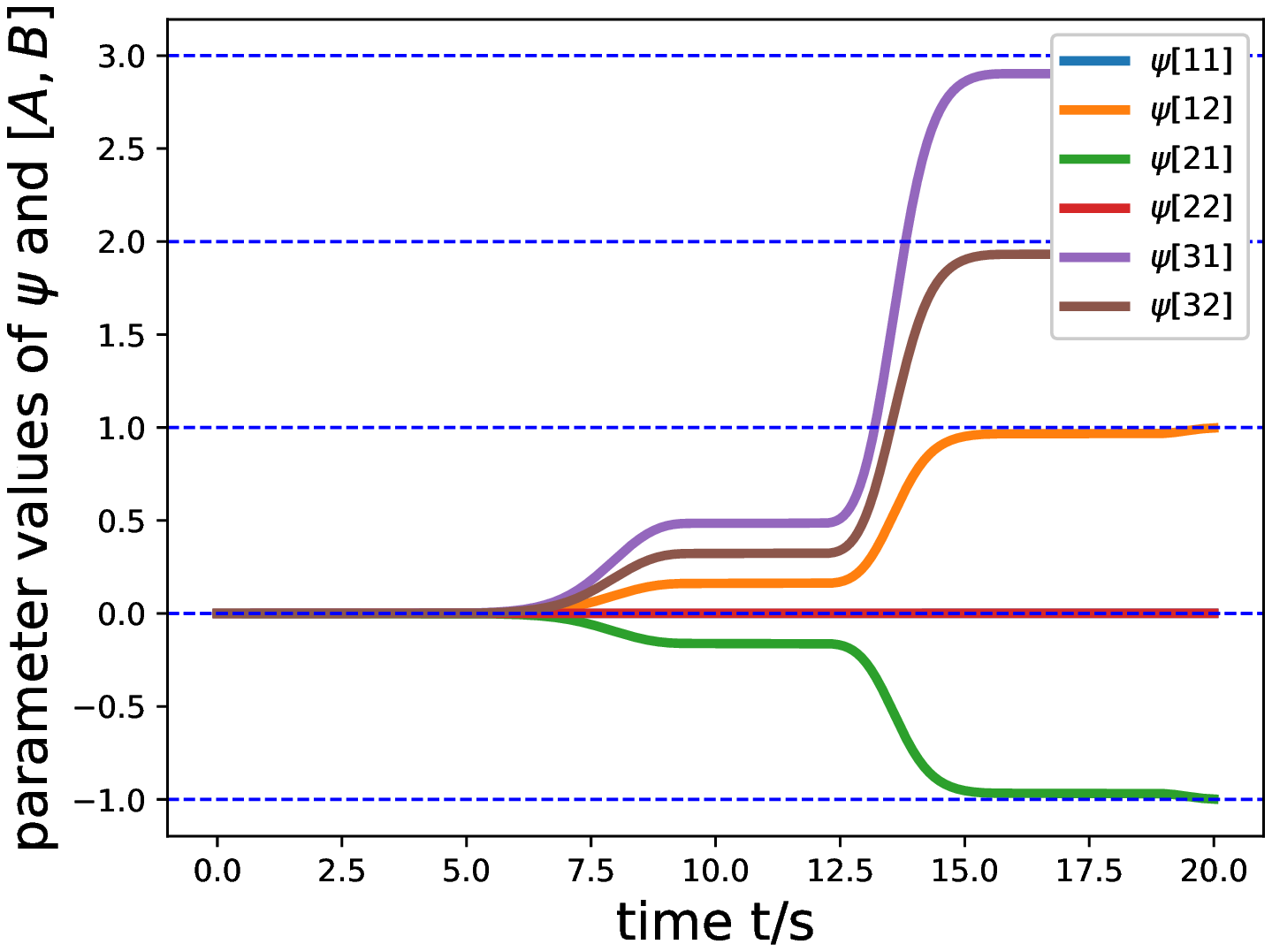}
		\end{minipage}
		\label{ex2_AB}
		}
	\subfigure[]{
		\begin{minipage}{4.1cm}
		\centering
		\includegraphics[width=4.1cm]{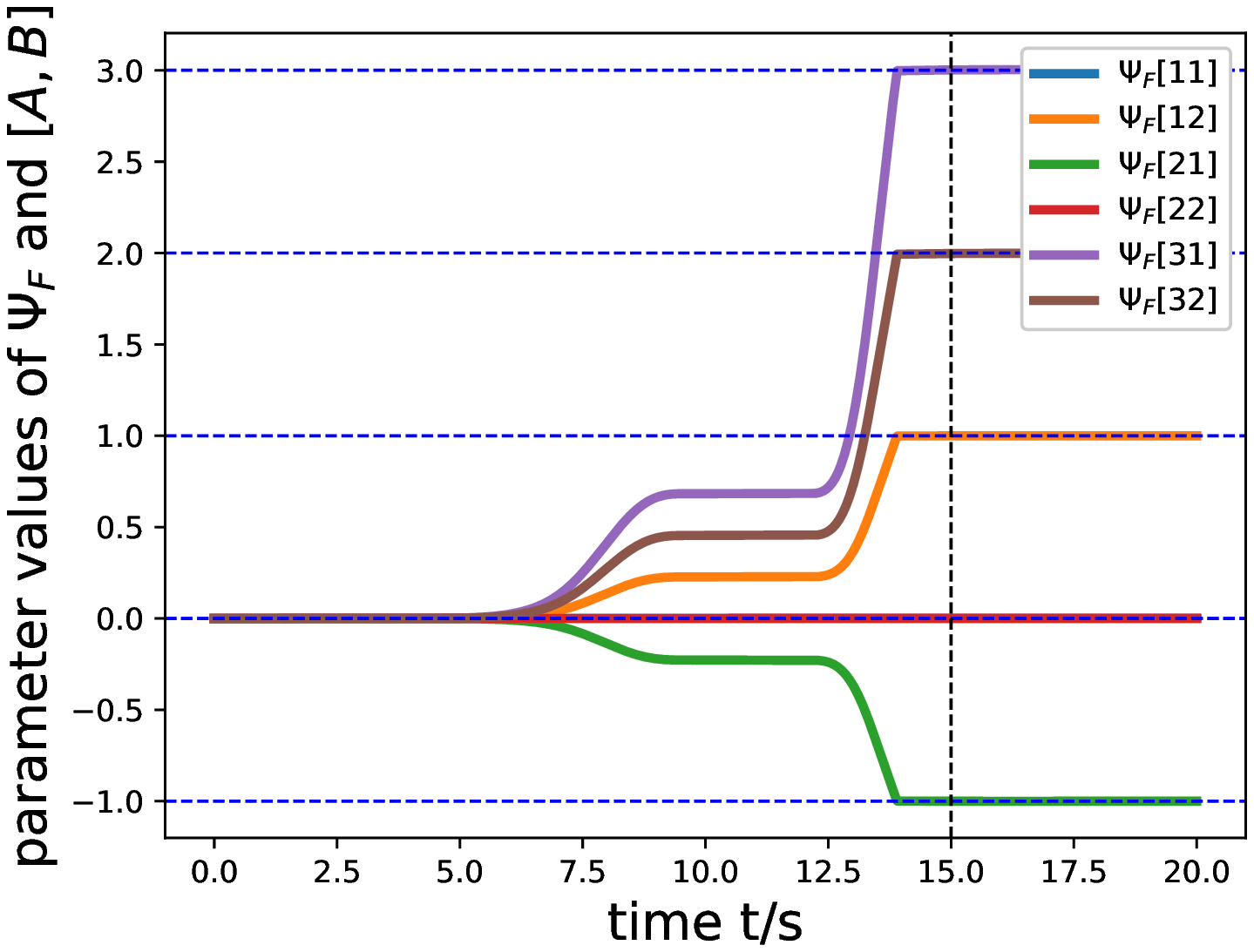}
		\end{minipage}
		\label{ex2_ABF}
		}
	\subfigure[]{
		\begin{minipage}{4.1cm}
		\centering
		\includegraphics[width=4.1cm]{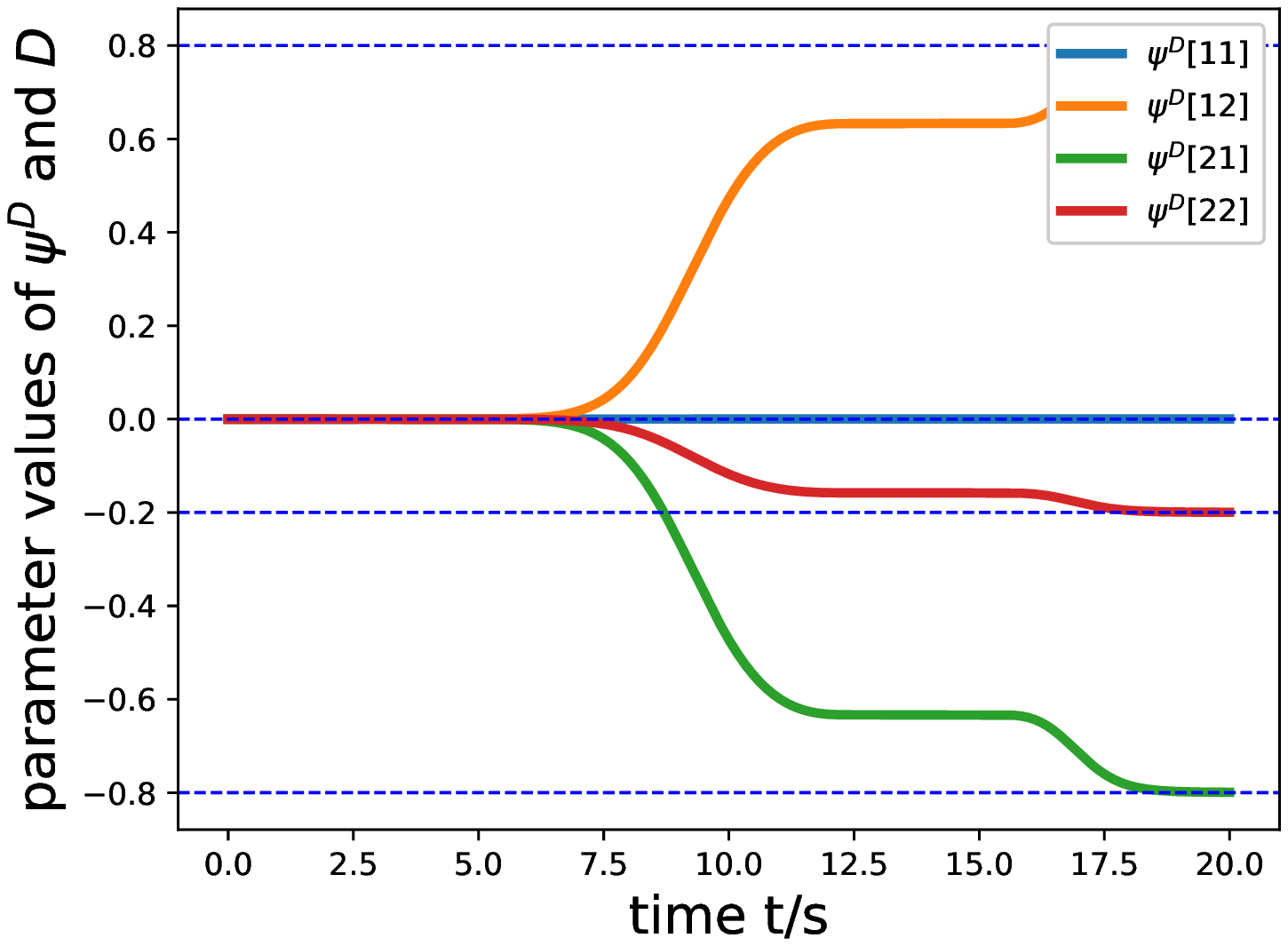}
		\end{minipage}
		\label{ex2_D}
		}
	\subfigure[]{
		\begin{minipage}{4.1cm}
		\centering
		\includegraphics[width=4.1cm]{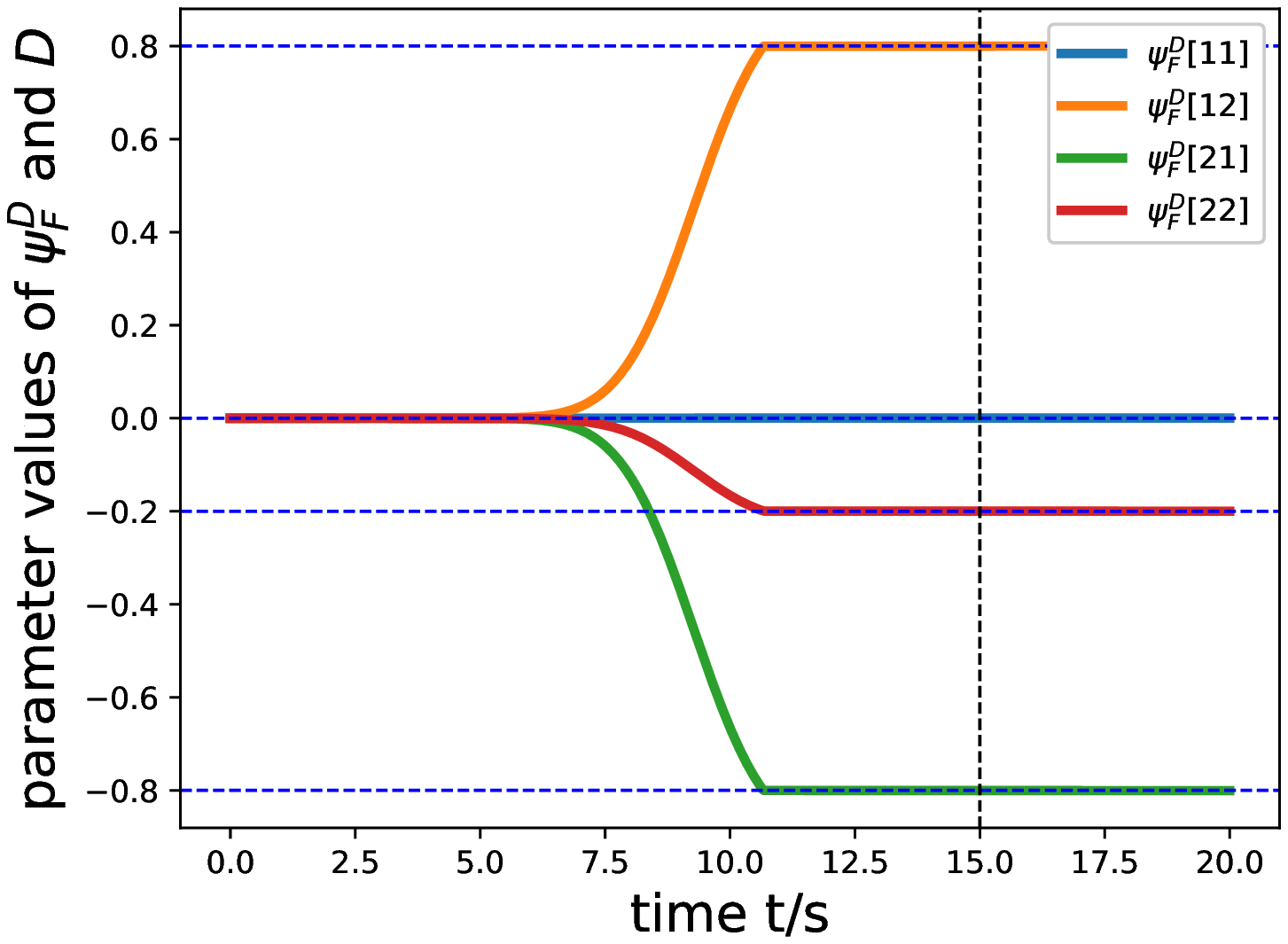}
		\end{minipage}
		\label{ex2_DF}
		}
		\caption{Signals' and estimated parameters' trajectories in Example \ref{ex2}. (a) Parameter estimation of $A$ and $B$ by $\psi$ under tuning law \eqref{linear_regression_updatelaw}. (b) Parameter estimation of $A$ and $B$ by $\psi_F$ under tuning law \eqref{intergrated_finiteestimation}. (c) Parameter estimation of $D$ by $\psi^D$ under tuning law \eqref{linear_regression_updatelaw}. (d) Parameter estimation of $D$ by $\psi^D_F$ under tuning law \eqref{intergrated_finiteestimation}.}
		\label{ex2_parameterestimation}
\end{figure}

\begin{figure}[t]
	\centering
	\subfigure[]{
		\begin{minipage}{8.1cm}
		\centering
		\includegraphics[width=8.1cm]{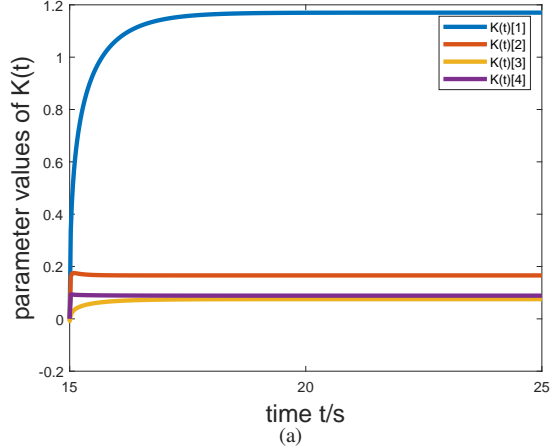}
		\end{minipage}
		\label{ex2_Kt}
		}
	\subfigure[]{
		\begin{minipage}{8.1cm}
		\centering
		\includegraphics[width=8.1cm]{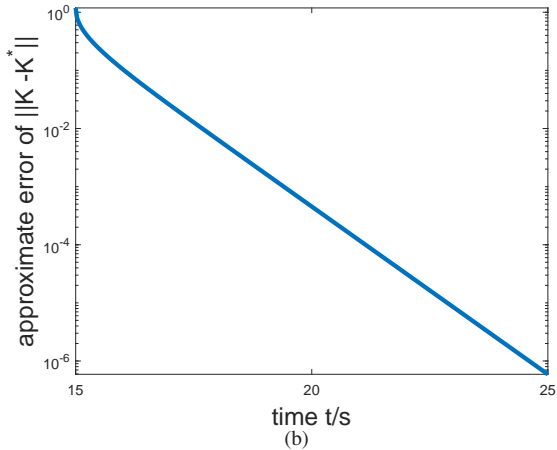}
		\end{minipage}
		\label{ex2_Kerror}
		}
		\caption{approximation trajectories of control gain in Example \ref{ex2}. (a) Evolutionary trajectories of searching based gain matrix $K(t)$ with exponential convergence. (b) Error of $\norm{K(t)-K^*}$ where $K^*$ is computed from LQR toolbox, which implies the optimality of the algorithm.}
		\label{ex2_optimalgains}
\end{figure}

\begin{figure}[t]
	\centering
	\subfigure[]{
		\begin{minipage}{8.1cm}
		\centering
		\includegraphics[width=8.1cm]{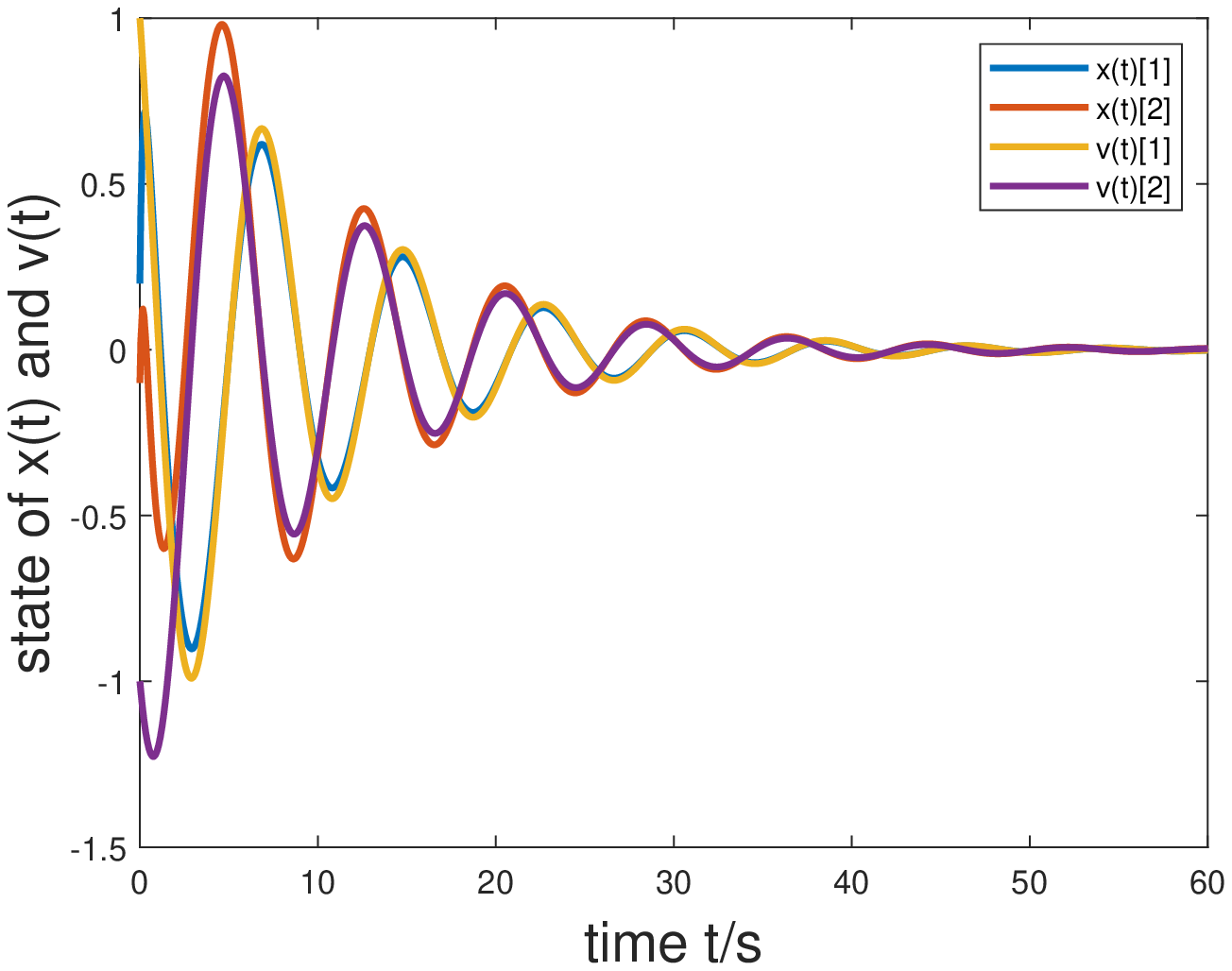}
		\end{minipage}
		\label{ex2_xvt}
		}
	\subfigure[]{
		\begin{minipage}{8.1cm}
		\centering
		\includegraphics[width=8.1cm]{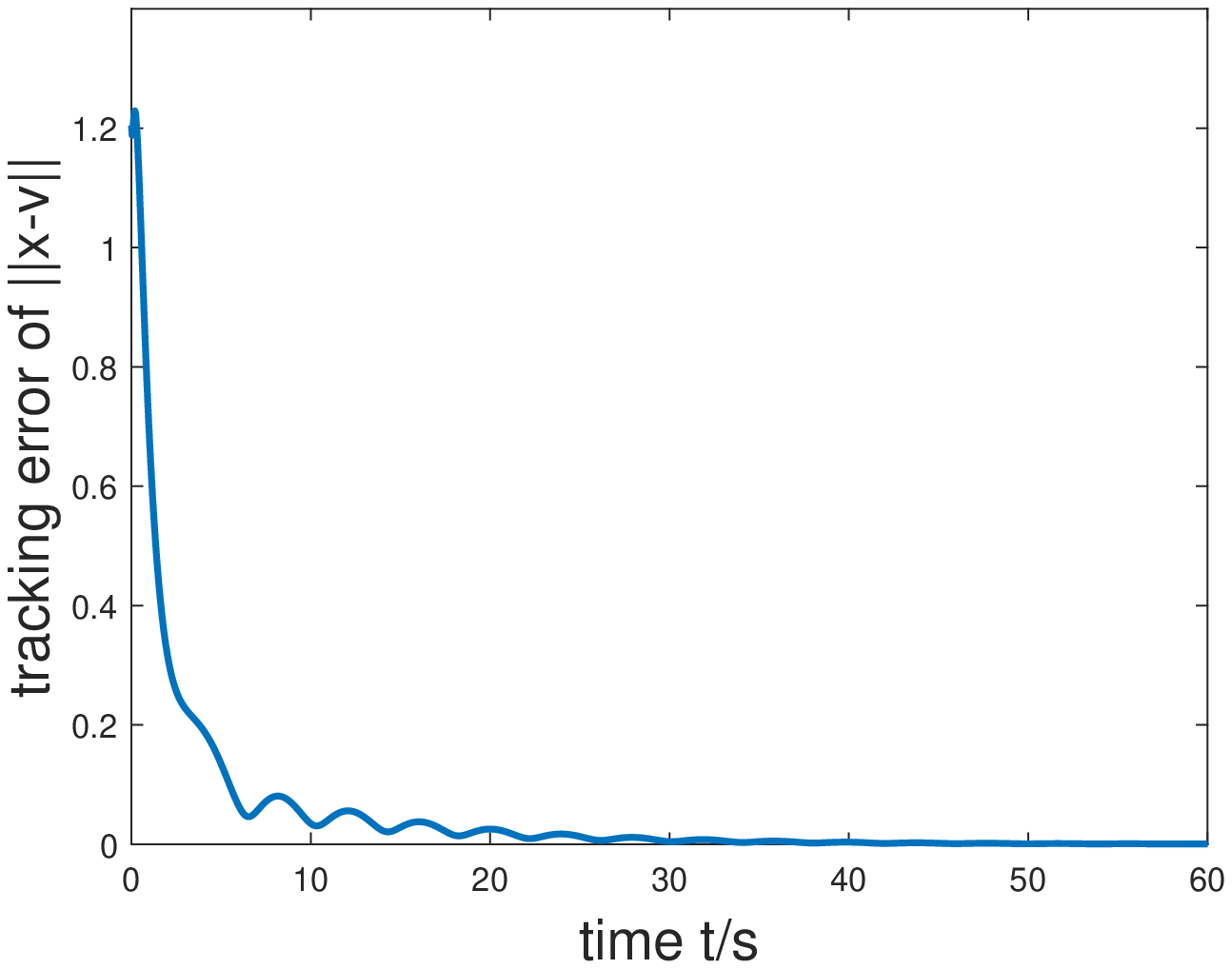}
		\end{minipage}
		\label{ex2_xverror}
		}
		\caption{Tracking trajectories of leader-follower systems with $u = -K(t)X$ in Example \ref{ex2}. (a) Trajectories of $x(t)$ and $v(t)$ with $x_0 = [0.2 -0.1]^\Tran$ and $v_0 = [1.0 -1.0]^\Tran$. (b) Error of $\norm{x(t) - v(t)}$ reflecting that the tracking objective is achieved.}
		\label{ex2_trackingtraj}
\end{figure}

Two simulation experiments are presented in this section to show the the effectiveness of parameter estimation and optimal solution searching, respectively.

\begin{example}
	\label{ex1}
For an uncoupled LCL inverter-based generation system defined in \cite{example_generator}, its dynamics is given as
\begin{equation}
	\dot x(t) = \left[\begin{array}{lll}
				-\frac{R_1}{L_1} & -\frac{1}{L_1} & 0\\
				\frac{1}{C} & 0 & -\frac{1}{C} \\
				0 & \frac{1}{L_2} & -\frac{R_2}{L_2}
			 \end{array}\right] x(t) + \left[\begin{array}{l}
				 \frac{1}{L_1}\\
				 0\\
				 0
			 \end{array}\right] u(t)
\end{equation}
in which $x(t) = [I_L,V_C,I_O]^\Tran$ and $u(t) = V_I$. Without loss of generality, the parameter values are modified for simplicity. The values of electronic components are chosen as $R_1 = 0.2\Omega$, $R_2 = 0.5\Omega$, $L_1 = 25mH$, $L_2 = 55mH$ and $C = 0.3F$. When the values of some components are unknown but the state $x(t)$ and input $u(t)$ are measurable, one can use Step 1 in Algorithm \ref{alg_integrated} to estimate these values.
\par The initial state $x(0) = [0,0,0]^\Tran$. Set the input voltage as $u(t) =100  sin(10t)V$, $t\in[0,25s]$ to force system working and collect the data of $x(t)$ and $u(t)$, while computing $s_0(t)$ as \eqref{aux_signal}. According to \eqref{linear_regression}, the parameters of designed filters is $\lambda = 0.1$, $\lambda_1 = 0.001$, $\lambda_2 = 0.01$, $\lambda_2 = 0.1$ and $\lambda_4 = 0.5$. Following the Algorithm \ref{alg_integrated}, the estimation matrix $\Psi\in\Real^{4\times 3}$ with initial value $0$ of all elements takes the learning law \eqref{linear_regression_updatelaw} with $\alpha = 0.05$. It is shown in Fig. \ref{ex1_fig_delta_1} that the parameters are not converge to the exact solution. Then, the evolutionary trajectory of $\Delta(t)$ is shown in Fig. \ref{ex1_fig_delta_2}, where $\Delta(t)\in \mathcal{L}_2$ that is contrary to the assumptions in Proposition \ref{proposition_1}. Therefore, the convergence under \eqref{linear_regression_updatelaw} may not be ensured. Note that PE condition is not satisfied either in this situation. Since IE condition is satisfied, it is feasible to introduce finite-time algorithm \eqref{finitetime_law} to solve the problem. According to Fig. \ref{ex1_fig_delta_2}, let the learning interval be $(0,25s)$, it is observed that $\int_{0s}^{25s} \Delta^2(\tau)d\tau \ge 40$. It is verified that IE condition \eqref{IEcondition_ln} holds when setting $\sigma = 0.60$. It is shown in Fig. \ref{ex1_fig_delta_3} that the estimation matrix $\Psi$ converges to the exact solution within $25s$, which illustrates the effectiveness. The norm-based error is shown in Fig. \ref{ex1_fig_delta_4}. Last but not least, both algorithm achieves element-wise approximation, i.e. each element is approximated asymptotically, as analyzed in \eqref{regression_element}.
\end{example}

\begin{example}
	\label{ex2}
	A controlled system is described by
	\begin{align}
		\dot x =& \left[\begin{array}{ll}
			0 & 1.0\\
			-1.0 & 0
		 \end{array}\right]x + \left[\begin{array}{llll}
			3.0 \\
			2.0
		 \end{array}\right]u,\\
		\dot v =& \left[\begin{array}{ll}
			0.0 & 0.8\\
			-0.8 & -0.2
		 \end{array}\right] v, 
	\end{align}
	where all of the parameter matrices are unknown. The control target is to design the controller $u$ with measurable state $x$ and $v$ to minimize the tracking performance \eqref{LQR_performance} with $\gamma = 0.5$, $Q = I$ and $R = 1$. With the ARE \eqref{ARE_P} including the discounted factor, the numerical solution of $P^*$ and optimal control gains $K^*$ can be obtained via LQR toolbox as
	\begin{align*}
		P^* =& \left[\begin{array}{llll}
			0.4645  & -0.2166  &  0.0459  &  0.0170\\
   			-0.2166 &   0.4543 &  -0.0524 &   0.0052\\
    		0.0459  & -0.0524  &  0.8389  &  0.0607\\
    		0.0170  &  0.0052  &  0.0607  &  0.7348\\
		 \end{array}\right],\\
		 K^* =& \left[\begin{array}{llll}
			1.1701266  &  0.1659746  &  0.0753407  &  0.0886391
		 \end{array}\right].
	\end{align*}
	Noticing that this method only works when the parameter matrices of system dynamics are known exactly and the toolbox is callable. When these conditions are hard to be met,
	the optimal control input can be approximated using Algorithm \ref{alg_integrated}.
	\par For simplicity, the settings of filters, control inputs and learning rate are the same as Example \ref{ex1}. In this way, the systems can be sufficiently excited and satisfied IE condition. Let $\sigma = 0.71$, the parameter estimation of $A$, $B$ and $D$ are approximated by $\psi$ and $\psi^D$ initialized by zero matrices, which is shown in Fig. \ref{ex2_AB} and Fig. \ref{ex2_D}. The finite-time estimation by $\psi_F$ and $\psi^D_F$ is shown in Fig. \ref{ex2_ABF} and Fig. \ref{ex2_DF} which implies that the estimation is done within $15s$.
	\par Because the parameter matrices are estimated and known after $15s$, the optimal solution can be obtained by searching based method, i.e. Step 2 in Algorithm \ref{alg_integrated}. Let $K_0=[0, 0, 0, 0]$. The eigenvalues of $(\mathcal{A} - \mathcal{B}K_0) - 0.25I$ are $(-0.25,i)$, $(-0.25,-i)$, $(-0.35,0.7937i)$ and $(-0.35,-0.7937i)$ verifying that $K_0 \in S_0$. Updating $K(t)$ with \eqref{search_alg}, the evolutionary trajectories of each element in $K(t)$ and error of $K(t) - K^*$ are shown in Fig. \ref{ex2_optimalgains}. Within $10s$, the control gain $K(t)$ approximates the numerical solution of ARE \eqref{ARE_P} and 
	\begin{align*}
		 K(25) =& [\begin{array}{llll}
			1.170126  &  0.1659746 &  0.0753407  &  0.0886391
		 \end{array}].
	\end{align*}
	\par With this approximated solution, the optimal controller can be designed by $u(t) = -K(25) X(t)$ where $X(t) = col(x(t)-v(t),v(t))$. The tracking trajectories can be seen in Fig. \ref{ex2_trackingtraj} which implies that two heterogeneous systems achieve the goal of state. Moreover, it is seen from \ref{ex2_Kerror} that the gradient-based method can approximate the optimal solutions with exponential convergence rate, which supports the result in Theorem \ref{theorem2}.
\end{example}
\section{Conclusion}\label{Section_conclusion}
This paper has investigated the optimal tracking control problem for unknown heterogeneous linear systems. The parameter estimation has been proposed first using measurable state and input, proving the finite-time accurate estimation under modified IE condition. With the estimated matrices of system dynamics, the optimal controller can be solved from a new ARE of augmented systems with discounted factor. A gradient descent method has been utilized to approximate the optimal solutions. It has been proven that the gradient-based method with appropriate initialization would result in exponential convergence rate and optimality of the numerical solutions. Last but not least, all of the conditions are online verifiable.
\par On the one hand, the robustness of presented method needs further analysis because the time delay,  unmeasurable disturbances and uncertain terms always exist in real systems. On the other hand, the validation of gradient method under more complex conditions such as sparse constraint remains an open topic.

{	\footnotesize
\bibliographystyle{IEEEtran}
\bibliography{mybib}

\begin{thebibliography}{10}
\providecommand{\url}[1]{#1}
\csname url@samestyle\endcsname
\providecommand{\newblock}{\relax}
\providecommand{\bibinfo}[2]{#2}
\providecommand{\BIBentrySTDinterwordspacing}{\spaceskip=0pt\relax}
\providecommand{\BIBentryALTinterwordstretchfactor}{4}
\providecommand{\BIBentryALTinterwordspacing}{\spaceskip=\fontdimen2\font plus
\BIBentryALTinterwordstretchfactor\fontdimen3\font minus
  \fontdimen4\font\relax}
\providecommand{\BIBforeignlanguage}[2]{{%
\expandafter\ifx\csname l@#1\endcsname\relax
\typeout{** WARNING: IEEEtran.bst: No hyphenation pattern has been}%
\typeout{** loaded for the language `#1'. Using the pattern for}%
\typeout{** the default language instead.}%
\else
\language=\csname l@#1\endcsname
\fi
#2}}
\providecommand{\BIBdecl}{\relax}
\BIBdecl

\bibitem{optimal_linear_books}
H.~Kwakernaak and R.~Sivan, \emph{Linear optimal control systems}.\hskip 1em
  plus 0.5em minus 0.4em\relax Wiley-interscience New York, 1972, vol.~1.

\bibitem{optimal_linear_quadratic_books}
B.~D. Anderson and J.~B. Moore, \emph{Optimal control: linear quadratic
  methods}.\hskip 1em plus 0.5em minus 0.4em\relax Courier Corporation, 2007.

\bibitem{LQRHtracking_unknownnonlinear_Jiang}
H.~Modares, F.~L. Lewis, and Z.-P. Jiang, ``H$_\infty$ tracking control of
  completely unknown continuous-time systems via off-policy reinforcement
  learning,'' \emph{IEEE Transactions on Neural Networks and Learning Systems},
  vol.~26, no.~10, pp. 2550--2562, 2015.

\bibitem{mu_rubust_tracking_unmatched}
C.~Mu, Y.~Zhang, Z.~Gao, and C.~Sun, ``Adp-based robust tracking control for a
  class of nonlinear systems with unmatched uncertainties,'' \emph{IEEE
  Transactions on Systems, Man, and Cybernetics: Systems}, vol.~50, no.~11, pp.
  4056--4067, 2020.

\bibitem{mu_ADP_datadriven}
C.~Mu, D.~Wang, and H.~He, ``Data-driven finite-horizon approximate optimal
  control for discrete-time nonlinear systems using iterative hdp approach,''
  \emph{IEEE Transactions on Cybernetics}, vol.~48, no.~10, pp. 2948--2961,
  2018.

\bibitem{shnbo_PETC}
S.~Wang, S.~Wen, K.~Shi, X.~Zhou, and T.~Huang, ``Approximate optimal control
  for nonlinear systems with periodic event-triggered mechanism,'' \emph{IEEE
  Transactions on Neural Networks and Learning Systems}, pp. 1--10, 2021.

\bibitem{Universal_approximation}
K.~Hornik, M.~Stinchcombe, and H.~White, ``Universal approximation of an
  unknown mapping and its derivatives using multilayer feedforward networks,''
  \emph{Neural Networks}, vol.~3, no.~5, pp. 551--560, 1990.

\bibitem{LQRtracking_discounted_nonlinear_Luo}
B.~Luo, D.~Liu, T.~Huang, and D.~Wang, ``Model-free optimal tracking control
  via critic-only q-learning,'' \emph{IEEE Transactions on Neural Networks and
  Learning Systems}, vol.~27, no.~10, pp. 2134--2144, 2016.

\bibitem{LQRtracking_two_timescale_discounted}
W.~Xue, J.~Fan, V.~G. Lopez, Y.~Jiang, T.~Chai, and F.~L. Lewis, ``Off-policy
  reinforcement learning for tracking in continuous-time systems on two time
  scales,'' \emph{IEEE Transactions on Neural Networks and Learning Systems},
  pp. 1--13, 2020.

\bibitem{LQR_unmeasurableD}
S.~A.~A. Rizvi, A.~J. Pertzborn, and Z.~Lin, ``Reinforcement learning based
  optimal tracking control under unmeasurable disturbances with application to
  hvac systems,'' \emph{IEEE Transactions on Neural Networks and Learning
  Systems}, pp. 1--11, 2021.

\bibitem{linear_optimal_unknown_state_initial}
Y.~Jiang and Z.-P. Jiang, ``Computational adaptive optimal control for
  continuous-time linear systems with completely unknown dynamics,''
  \emph{Automatica}, vol.~48, no.~10, pp. 2699--2704, 2012.

\bibitem{initialexcitation_comparetorank}
S.~K. Jha, S.~B. Roy, and S.~Bhasin, ``Initial excitation-based iterative
  algorithm for approximate optimal control of completely unknown lti
  systems,'' \emph{IEEE Transactions on Automatic Control}, vol.~64, no.~12,
  pp. 5230--5237, 2019.

\bibitem{RL_concurrent_tracking_converge}
C.~Chen, H.~Modares, K.~Xie, F.~L. Lewis, Y.~Wan, and S.~Xie, ``Reinforcement
  learning-based adaptive optimal exponential tracking control of linear
  systems with unknown dynamics,'' \emph{IEEE Transactions on Automatic
  Control}, vol.~64, no.~11, pp. 4423--4438, 2019.

\bibitem{track_IRL}
H.~Modares and F.~L. Lewis, ``Linear quadratic tracking control of
  partially-unknown continuous-time systems using reinforcement learning,''
  \emph{IEEE Transactions on Automatic Control}, vol.~59, no.~11, pp.
  3051--3056, 2014.

\bibitem{Annual_Reviews}
R.~Ortega, V.~Nikiforov, and D.~Gerasimov, ``On modified parameter estimators
  for identification and adaptive control. a unified framework and some new
  schemes,'' \emph{Annual Reviews in Control}, vol.~50, pp. 278--293, 2020.

\bibitem{Rushikesh_tac_concurrent_pestimation}
R.~Kamalapurkar, B.~Reish, G.~Chowdhary, and W.~E. Dixon, ``Concurrent learning
  for parameter estimation using dynamic state-derivative estimators,''
  \emph{IEEE Transactions on Automatic Control}, vol.~62, no.~7, pp.
  3594--3601, 2017.

\bibitem{PEcondition}
K.~S. Narendra and A.~M. Annaswamy, ``Persistent excitation in adaptive
  systems,'' \emph{International Journal of Control}, vol.~45, no.~1, pp.
  127--160, 1987.

\bibitem{IEcondition}
G.~Kreisselmeier and G.~Rietze-Augst, ``Richness and excitation on an
  interval-with application to continuous-time adaptive control,'' \emph{IEEE
  Transactions on Automatic Control}, vol.~35, no.~2, pp. 165--171, 1990.

\bibitem{IE_MRAC_2013}
G.~Chowdhary, T.~Yucelen, M.~Mühlegg, and E.~N. Johnson, ``Concurrent learning
  adaptive control of linear systems with exponentially convergent bounds,''
  \emph{International Journal of Adaptive Control and Signal Processing},
  vol.~27, no.~4, pp. 280--301, 2013.

\bibitem{robotic_finite_IE}
C.~Yang, Y.~Jiang, W.~He, J.~Na, Z.~Li, and B.~Xu, ``Adaptive parameter
  estimation and control design for robot manipulators with finite-time
  convergence,'' \emph{IEEE Transactions on Industrial Electronics}, vol.~65,
  no.~10, pp. 8112--8123, 2018.

\bibitem{DREM_begin}
S.~Aranovskiy, A.~Bobtsov, R.~Ortega, and A.~Pyrkin, ``Performance enhancement
  of parameter estimators via dynamic regressor extension and mixing*,''
  \emph{IEEE Transactions on Automatic Control}, vol.~62, no.~7, pp.
  3546--3550, 2017.

\bibitem{matrix_estimation_finite_time}
R.~Ortega, D.~N. Gerasimov, N.~E. Barabanov, and V.~O. Nikiforov, ``Adaptive
  control of linear multivariable systems using dynamic regressor extension and
  mixing estimators: Removing the high-frequency gain assumptions,''
  \emph{Automatica}, vol. 110, p. 108589, 2019.

\bibitem{finite_time_robust}
J.~Wang, D.~Efimov, and A.~A. Bobtsov, ``On robust parameter estimation in
  finite-time without persistence of excitation,'' \emph{IEEE Transactions on
  Automatic Control}, vol.~65, no.~4, pp. 1731--1738, 2020.

\bibitem{KleinDARE}
D.~Kleinman, ``On an iterative technique for riccati equation computations,''
  \emph{IEEE Transactions on Automatic Control}, vol.~13, no.~1, pp. 114--115,
  1968.

\bibitem{PI_ARE_Lee}
J.~Y. Lee, J.~B. Park, and Y.~H. Choi, ``On integral generalized policy
  iteration for continuous-time linear quadratic regulations,''
  \emph{Automatica}, vol.~50, no.~2, pp. 475--489, 2014.

\bibitem{QlearningWei}
Q.~Wei, D.~Liu, and G.~Shi, ``A novel dual iterative q-learning method for
  optimal battery management in smart residential environments,'' \emph{IEEE
  Transactions on Industrial Electronics}, vol.~62, no.~4, pp. 2509--2518,
  2015.

\bibitem{DRLQLW_review}
D.~Liu, S.~Xue, B.~Zhao, B.~Luo, and Q.~Wei, ``Adaptive dynamic programming for
  control: A survey and recent advances,'' \emph{IEEE Transactions on Systems,
  Man, and Cybernetics: Systems}, vol.~51, no.~1, pp. 142--160, 2021.

\bibitem{ARE_gradient_Lsmooth}
I.~Fatkhullin and B.~Polyak, ``Optimizing static linear feedback: Gradient
  method,'' 2020.

\bibitem{ARE_gradient_variable_CDC}
H.~Mohammadi, A.~Zare, M.~Soltanolkotabi, and M.~R. Jovanović, ``Global
  exponential convergence of gradient methods over the nonconvex landscape of
  the linear quadratic regulator,'' in \emph{2019 IEEE 58th Conference on
  Decision and Control (CDC)}, 2019, pp. 7474--7479.

\bibitem{model_freeLQR_TAC}
H.~Mohammadi, A.~Zare, M.~Soltanolkotabi, and M.~R. Jovanovic, ``Convergence
  and sample complexity of gradient methods for the model-free linear quadratic
  regulator problem,'' \emph{IEEE Transactions on Automatic Control}, pp. 1--1,
  2021.

\bibitem{1992_gradient_variable}
E.~Feron, V.~Balakrishnan, S.~Boyd, and L.~El~Ghaoui, ``Numerical methods for
  $h_2$ related problems,'' in \emph{1992 American Control Conference}, 1992,
  pp. 2921--2922.

\bibitem{convergence_lemma}
B.~Polyak, ``Gradient methods for the minimisation of functionals,'' \emph{USSR
  Computational Mathematics and Mathematical Physics}, vol.~3, no.~4, pp.
  864--878, 1963.

\bibitem{Lyapunov_equation}
C.-H. Lee, ``New results for the bounds of the solution for the continuous
  riccati and lyapunov equations,'' \emph{IEEE Transactions on Automatic
  Control}, vol.~42, no.~1, pp. 118--123, 1997.

\bibitem{adgrad}
J.~Duchi, E.~Hazan, and Y.~Singer, ``Adaptive subgradient methods for online
  learning and stochastic optimization.'' \emph{Journal of machine learning
  research}, vol.~12, no.~7, 2011.

\bibitem{example_generator}
K.~H. Ahmed, A.~M. Massoud, S.~J. Finney, and B.~W. Williams, ``A modified
  stationary reference frame-based predictive current control with zero
  steady-state error for lcl coupled inverter-based distributed generation
  systems,'' \emph{IEEE Transactions on Industrial Electronics}, vol.~58,
  no.~4, pp. 1359--1370, 2011.

\end{thebibliography}
}

\end{document}